\documentclass[sigconf]{acmart}

\AtBeginDocument{%
  }

\usepackage{soul}
\usepackage{color}
\usepackage{amsmath}

\usepackage{gensymb}
\usepackage{enumitem}

\usepackage{csquotes}

\newcommand{\systemName}{\textsc{XR-Objects}}
\newcommand{\paradigmName}{Augmented Object Intelligence}

\newcommand{\new}[1]{#1}
\copyrightyear{2024}
\acmYear{2024}
\setcopyright{rightsretained}
\acmConference[UIST '24]{The 37th Annual ACM Symposium on User Interface Software and Technology}{October 13--16, 2024}{Pittsburgh, PA, USA}
\acmBooktitle{The 37th Annual ACM Symposium on User Interface Software and Technology (UIST '24), October 13--16, 2024, Pittsburgh, PA, USA}
\acmDOI{10.1145/3654777.3676379}
\acmISBN{979-8-4007-0628-8/24/10}

\begin{document}

% \title{\systemName: Augmented Interaction for Analog Objects}

% \title{Augmented Object Intelligence: Making the Analog World Interactable with \systemName}

\title{Augmented Object Intelligence with \systemName}


\author{Mustafa Doga Dogan}
% \affiliation{%
%   \institution{MIT CSAIL, USA}}
\affiliation{%
  \institution{Google}
  \city{Zurich}
  \country{Switzerland}
  }
  \email{doga@dogadogan.com}

\author{Eric J. Gonzalez}
\affiliation{%
  \institution{Google}
  \city{Seattle, Washington}
  \country{USA}
  }
  \email{ejgonz@google.com}
  
  \author{Karan Ahuja}
% \affiliation{%
%   \institution{Northwestern University, USA}}
\affiliation{%
  \institution{Google}
  \city{Seattle, Washington}
  \country{USA}
  }
\email{karanahuja@google.com}
  
  \author{Ruofei Du}
\affiliation{%
  \institution{Google}
  \city{San Francisco, California}
  \country{USA}
  }
  \email{me@duruofei.com}
  
    \author{Andrea Colaço}
\affiliation{%
  \institution{Google}
  \city{Mountain View, California}
  \country{USA}
  }
  \email{andreacolaco@google.com}
  
    \author{Johnny Lee}
\affiliation{%
  \institution{Google}
  \city{Redmond, Washington}
  \country{USA}
  }
  \email{johnnylee@google.com}

    \author{Mar Gonzalez-Franco}
\affiliation{%
  \institution{Google}
  \city{Seattle, Washington}
  \country{USA}
  }
  \email{margon@google.com}
\authornote{Both authors share co-senior authorship.}

    \author{David Kim}
\affiliation{%
  \institution{Google}
  \city{Zurich}
  \country{Switzerland}
  }
  \email{kidavid@google.com}
\authornotemark[1]

\renewcommand{\shortauthors}{Dogan et al.}

\begin{abstract}

% \david{Proposing a title change to address 1ACs comment about toning down AOI. E.g. Like this or similar "XR-Objects: Towards World-Centric Interactions with Augmented Object Intelligence" to acknowledge that we don't invent AOI per se, and we are working towards that vision}

Seamless integration of physical objects as interactive digital entities remains a challenge for spatial computing. This paper explores \paradigmName~ (AOI) in the context of XR, an interaction paradigm that aims to blur the lines between digital and physical by equipping real-world objects with the ability to interact as if they were digital, where every object has the potential to serve as a portal to digital functionalities.
Our approach utilizes real-time object segmentation and classification, combined with the power of Multimodal Large Language Models (MLLMs), to facilitate these interactions without the need for object pre-registration.
We implement the AOI concept in the form of \systemName, an open-source prototype system that provides a platform for users to engage with their physical environment in contextually relevant ways using object-based context menus.
This system enables analog objects to not only convey information but also to initiate digital actions, such as querying for details or executing tasks.
Our contributions are threefold: (1) we define the AOI concept and detail its advantages over traditional AI assistants, (2) detail the \systemName~ system's open-source design and implementation, and (3) show its versatility through various use cases and a user study.

\end{abstract}






\begin{teaserfigure}
  \centering
  \includegraphics[width=1\textwidth]{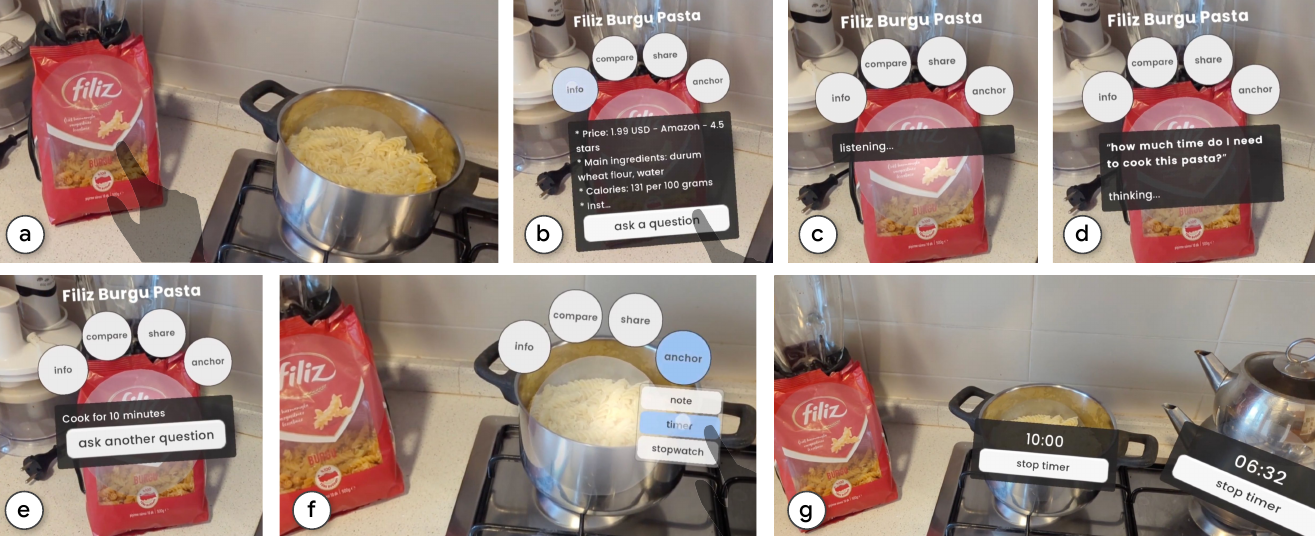}
  % https://docs.google.com/drawings/d/1_eKojSFtBeskouYJHC3HQZR_BydpbUokYi9H8_7de_0/edit?usp=drive_link
  \caption{\systemName~ allows users to (a) select and interact with real-world objects in AR as if they were digital objects. Automatically generated object-based AR context menus allow objects to (b) provide information about themselves, such as nutritional facts and ingredients. For example, a user (c, d, e) asks a question about the cooking time of pasta, and then (f, g) uses the answer to set a spatial timer widget anchored to the relevant pot in 3D space.
  % \ruofei{uses the answer to; for future work: is the timer manually added there or suggested to add there? e.g., after "how much time do I need to cook this pasta", instead of "ask another question", it could be "set a timer";} 
  % \ruofei{note that all a11y text are not enough, they are designed to be read by blind people and need to describe the figures. please do a thorough pass.} 
  }
  \Description{(a) A pack of pasta and a pot on a stove. (b) A menu panel overlaid on the pasta, which says: "Filiz Burgu Pasta. (c) The interface says "listening". (d) "how much time do I need to cook this pasta?" (e) Cook for 10 minutes" (f) The user taps "timer" (g) A timer on the pot, another one on the teapot.}
  \label{fig:teaser}
\end{teaserfigure}

\maketitle

\begin{figure*}[t!]
  \centering
  \includegraphics[width=0.84\linewidth]{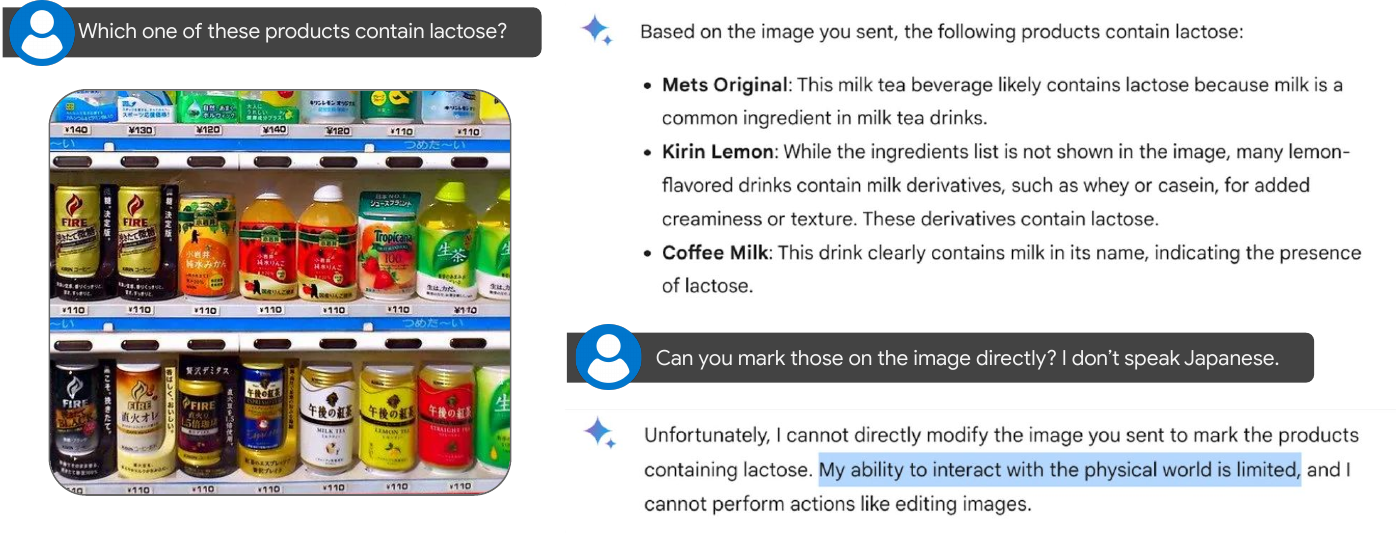}
  % https://docs.google.com/drawings/d/1LrYLkevHrLZ_wjz7oT-uM0dBH_4Hw4di1Ae9Tq0vF5g/edit?usp=drive_link
  \caption{Example of an interaction with a conventional multimodal AI assistant. While the model clearly has the capacity to produce reasonable scene understanding when an image and a prompt is provided as input, it fails in providing an anchored output that ties back to the original multimodal prompt.}
  \Description{Left: The user prompt says, "Which of these products contain lactose?" and there is a photo with several beverages. Right: The LLM gives a text answer with 3 product names. The user then asks: "Can you mark those on the image directly? I don't speak Japanese. The LLM responds: "Unfortunately, I cannot - my ability to interact with the physical world is limited.}
  \label{fig:StandardLLMInteraction}
\end{figure*}

\section{Introduction}

\new{Modern Extended Reality (XR) platforms come with a plethora of sensors, cameras, and advanced computer vision techniques to seamlessly blend virtual content with the physical world through color passthrough and scene understanding. However, despite these technological steps, the integration of real objects into the XR environment remains somewhat superficial, treating the physical world largely as a mere backdrop rather than an interactive component. In contrast, projects like \textit{RealityCheck}~\cite{hartmann_realitycheck_2019} or \textit{Remixed Reality}~\cite{lindlbauer_remixed_2018} show a future where digital and physical worlds could be closely intertwined together. Similarly, advancements in artificial intelligence (AI) are laying the groundwork for such a future, with breakthroughs in real-time unsupervised segmentation~\cite{tian_diffuse_2023} combined with in-painting~\cite{xiang_deep_2023} or generative AI content generation~\cite{kerbl_3d_2023}.}

\new{The wide availability of machine learning (ML) and computer vision technologies has also led to features that enhance digital interaction with the physical world at our fingertips. Tools like image-based search in \textit{Google Lens}\footnote{Google Lens: \url{https://lens.google}} and utility-focused Augmented Reality (AR) features in smartphones, such as text copy \& paste and real-time translation, are becoming increasingly common.
Together, these tools represent building blocks that could bring us closer to a future in which a total understanding of the world and its objects can be applied to our everyday interactions in XR.}

% \new{In this paper, we explore a novel interaction paradigm we term \textit{Augmented Object Intelligence (AOI)} that would allow any real object identified by the XR system to reveal digital data associated with the \textbf{analog object} and perform context-appropriate \textbf{digital actions} in a meaningful way.

\new{In this paper, we explore an interaction paradigm we term \textit{Augmented Object Intelligence (AOI)} in XR, which allows any \textbf{\textit{analog object}} identified by the XR system to reveal \textbf{\textit{digital data}} associated with it. This enables users to perform context-appropriate \textit{\textbf{digital actions} with respect to the object} in a meaningful way.
Our system \systemName~ embodies this idea and aims to demonstrate and investigate ``semantic equality" between real and virtual objects without the need for pre-registration.}

\new{Imagine a scenario as familiar \textbf{as right-clicking a digital file to open its context menu, but \textit{applied to physical objects within XR}} — such as \textit{right-clicking} on potatoes or pasta in a pot to start a cooking timer set to the correct duration, or filtering for gluten-free products on a grocery shelf through an XR interface (Figure~\ref{fig:teaser}).}

\new{The leap towards physical awareness in \systemName~ represents an advancement over traditional AR, which often relies on manual input or the use of physical tracking markers.
We uniquely combine developments in spatial understanding via technologies such as Simultaneous Localization and Mapping (SLAM), available in \textit{ARCore}~\cite{google_arcore_2024} and \textit{ARKit}~\cite{inc_arkit_2024}, and machine learning models for object segmentation and classification (\textit{COCO}~\cite{lin_microsoft_2015} via \textit{MediaPipe}~\cite{lugaresi_mediapipe_2019}). These technologies enable us to implement \textbf{object instance-based AR interactions} with semantic depth and achieve \textbf{live detection and 3D localization without pre-registration}.
We also integrate a Multimodal Large Language Model (MLLM) %, \textit{Google Gemini}~\cite{team2023gemini},
into our system, which further enhances our ability to automate the recognition of objects and their specific semantic information within XR spaces.}

\vspace{0.2cm}
Our contributions are threefold:

\vspace{-0.1cm}
\begin{itemize}[leftmargin=0.5cm]
    \item We introduce the concept of \textbf{\paradigmName{}} (AOI) for XR, a paradigm shift towards seamless integration of real and virtual content in XR using AI and object-based context menu interfaces.
    \item We detail the open-sourced\footnote{\new{\systemName~ open-source project: \url{https://github.com/google/xr-objects}}}  design and implementation of \textbf{\systemName}, our prototypical system that exemplifies AOI, alongside an exploration of diverse use cases to demonstrate its potential.
    \item We provide a comparative evaluation between standard prompt-based LLM interfaces and our AOI approach for contextual information retrieval and object-centric interaction in AR, highlighting AOI's significant reduction in task completion time and its enhanced ease of use and satisfaction.     
\end{itemize}

\new{We demonstrate our prototype that integrates these AR and AI components in a seamless way, which is implemented for smartphones to provide access to a broad audience, as commercial XR headsets with large field of view (e.g., \textit{Meta Quest 3}) do not yet give programmatic access to the user’s camera stream.
% By open-sourcing XR-Objects, we invite the research community to engage with our work to help pave the way for a future where digital and physical realities merge seamlessly, and hope that this work will be enabled on commercial headsets soon.
%This vision of Programmable Reality represents not just a technological advancement, but a new chapter in how users interact with the world around them, making everyday tasks more intuitive and integrated into our digital lives.
% Our approach advances AR by integrating a spatial and object instance-based interaction framework with live detection and 3D localization without pre-registration.
By \textbf{open-sourcing our system}, we aim to foster further innovation in the field, ultimately bringing us closer to a future where the physical and digital realms interact seamlessly.}

\section{Related Work}
This section provides an overview of previous work in % the realms of
user interface (UI) design principles, blended reality interactions, and advancements in AR technology~\cite{speicher_what_2019}. These areas form the foundation upon which our research builds, which aims to enhance the interaction between users and the physical world through \systemName.

%our work draws upon and integrates research from various areas of HCI, AR, and AI and focuses on making object-specific content applicable to all objects, including non-electronic and markerless (analog) objects. We aim to create a more seamless way for users to interact with the physical world in XR.

\subsection{Fundamentals of UI}
The challenge of bridging the cognitive gap between human users and computational systems has been a central theme in HCI. Traditional approaches have employed various layers of representations to mediate this interaction, manifesting most recognizably in the UIs of devices ranging from PCs and smartphones to IoT devices and automotive systems~\cite{tidwell_designing_2020}. Despite these advancements, the resurgence of command-line interface (CLIs) in contemporary AI interactions, as seen in the usage of prompts with large language models (LLMs) \cite{suh_luminate_2024}, suggests a potential oversimplification of user interaction paradigms. This regression shows the necessity of reevaluating our approach to UI design in the age of AI and spatial computing~\cite{suzuki_xr_2023}, where the intricacy of human intentions and the computational interpretation thereof demand a more nuanced form of representation.  Figure~\ref{fig:StandardLLMInteraction} demonstrates that the multimodal AI assistant clearly has the capacity to produce reasonable scene understanding when an image and a prompt is provided as input, but it fails in providing a robustly anchored output or interaction capabilities that tie back to the original multimodal prompt.

Significantly, the role of UIs extends beyond mere facilitation of interaction, which shapes the user’s ability to navigate, understand, and command the computational system. Effective UIs support essential cognitive functions such as memory, discovery, and articulation \cite{blackwell_reification_2006}, thus becoming a key factor in the widespread adoption and utility of AI and spatial computing technologies.

In addressing these challenges, our work revisits the utility of \textbf{\textit{context menus}} — a familiar paradigm in desktop computing \cite{airth_navigation_1993, zeng_multiple_2014, banovic_design_2011} — and explores their potential in fostering familiar interactions with physical objects within XR environments.
Previous demonstrations~\cite{lepinski_context_2009} showed the potential of such ubiquitous context menus through simulated digital notes via projection. In this work, we show a fully functional prototype using AR.

\subsection{Extended Reality}
XR systems represent a rapidly developing field within HCI, with recent works aiming to make the boundaries between real and virtual environments indistinguishable \cite{lindlbauer_combining_2016, dogan_fabricate_2022}. While a comprehensive literature review~\cite{auda_scoping_2023} is beyond the scope of this section, we highlight key themes from the literature that inspired our research.

\new{\textit{Interacting with Real-World Affordances}:
 As XR platforms become increasingly accessible, enhancing physical-aware interactions can significantly elevate user experiences. \textit{DepthLab}~\cite{du_depthlab_2020} exemplifies advancements in real-time 3D interactions with the real world using depth maps.
\textit{InteractionAdapt} \cite{cheng_interactionadapt_2023} optimizes VR work-spaces for efficient interaction across diverse physical environments.}

\new{\textit{Manipulating Perception in XR}: Gonzalez-Franco and Lanier~\cite{gonzalez-franco_model_2017} explored how perceptual manipulations can be effectively modeled within VR environments to enrich user experience. 
Bonnail et al. explore how XR can leverage human memory limitations to influence user perception and behavior~\cite{bonnail_memory_2023}. Concurrently, \textit{causality-preserving asynchronous reality} enables users to interact with events in a causally accurate manner despite temporal delays~\cite{fender_causality-preserving_2022}.
Tseng et al. highlighted the risks associated with such manipulations~\cite{tseng_dark_2022} and proposed mitigations against their malicious use~\cite{tseng_understanding_2023}.}

% \textit{Prototyping XR Systems}:
% In \textit{VRception}, Gruenefeld et al. discuss a methodology for rapid prototyping of cross-reality systems within VR settings \cite{gruenefeld2022vrception}. This approach aligns with our goals in \systemName, aiming to reduce developmental barriers and enable swift knowledge transfer across multiple realities.

\new{\textit{Balancing Immersion and External Awareness}: 
Critical to XR systems is managing the balance between immersion in the virtual world and awareness of the real world. Studies such as those by Kudo et al. emphasize the importance of smoothly transitioning users and devices between realities, while also maintaining bystander awareness~\cite{kudo_towards_2021}. Further guidelines on this balance are discussed in works by Gonzalez-Franco et al. for harmonizing user experience across dimensions \cite{gonzalez-franco_guidelines_2024, gonzalez_xdtk_2024}.}

%\subsection{Detecting and Interacting with AR Objects}
\subsection{Interacting with Physical Objects in XR}
Despite the considerable advances in AI’s capability to generate and understand complex content, our physical world remains predominantly analog, with only a fraction of daily activities and tools being enhanced by digital technology~\cite{rogers_understanding_2020, wang_texturesight_2024, dogan_sensicut_2021}. This analog nature of human experiences, from basic needs fulfillment to complex task execution, presents a significant challenge in integrating digital intelligence in a manner that feels both natural and easy to grasp.

\begin{figure}[h]
  \centering
  \includegraphics[width=0.8\linewidth]{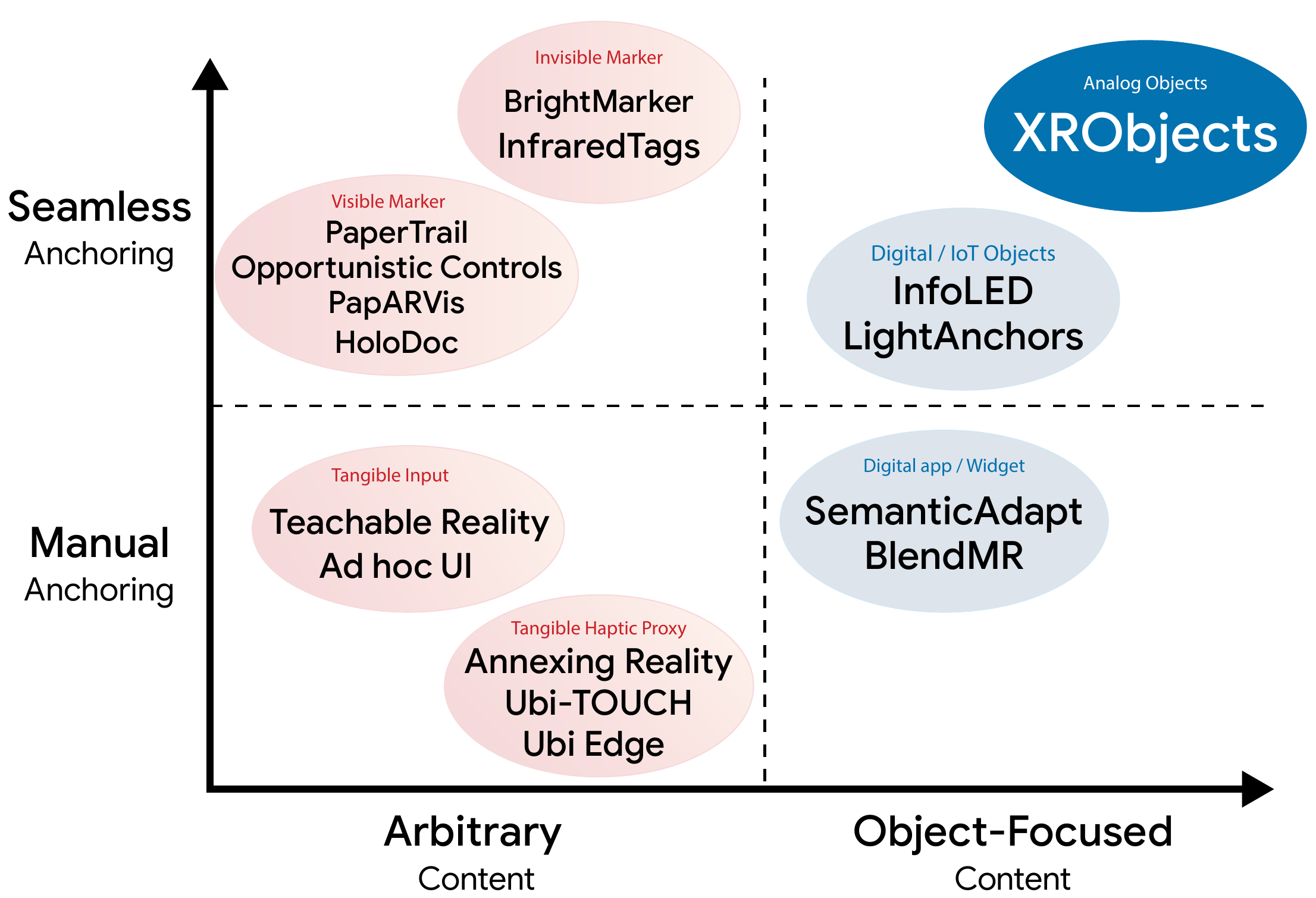}
  % https://docs.google.com/drawings/d/1qLiHIO98pUuprqJ0TgjT1MaLD4ZcxxJQZSU0Px2eq70/edit?usp=drive_link
  \caption{The landscape of physical object interactions in XR classified across two dimensions: anchoring and content.
  %\ruofei{larger font or larger width of the figure}
  }
  \Description{A diagram with two axes: The x-axis says "arbitrary content vs. object-focused content". The y-axis: "manual anchoring" vs "seamless anchoring". There are related work groups in this 2D space, including visible marker, invisible marker, tangible input, tangible haptic proxy, digital app, digital objects.}
  \label{fig:RelatedWork}
\end{figure}

Several research efforts have explored ways to bridge this gap by leveraging XR technologies to enable interaction with real-world objects~\cite{xu_xair_2023}. Figure~\ref{fig:RelatedWork} shows the landscape of physical object interactions in XR classified across two dimensions: anchoring (manual vs. seamless) and content (arbitrary vs. object-focused).

\new{Building on the foundational work in tangible bits~\cite{ishii_tangible_1997, ishii_tangible_2008}, some approaches rely on pre-registration of objects or manual setups to achieve tangible input} \cite{du_opportunistic_2022, monteiro_teachable_2023, zhu_mecharspace_2022, campos_zamora_moirewidgets_2024} or tangible haptic proxy \cite{he_ubi_2023, jain_ubi-touch_2023, hettiarachchi_annexing_2016}.
In contrast to these manual processes, other works utilized markers to more automatically execute the object detection and AR content anchoring~\cite{dogan_ubiquitous_2024}. For instance, researchers embedded visible \cite{rajaram_paper_2022, henderson_opportunistic_2008, chen_augmenting_2020, li_holodoc_2019} or invisible markers \cite{dogan_infraredtags_2022, dogan_brightmarker_2023, dogan_standarone_2023, dogan_g-id_2020, dogan_demonstrating_2022} to documents and 3D objects for AR purposes. However, these methods often impose limitations on the types of objects that can be interacted with, requiring specific fabrication or preparation.

Our exploration acknowledges the complexity of translating digital interactions to physical objects and aims to bridge this divide by enhancing any physical object with digital functionality and contextual interaction capabilities.

Once the objects have been identified and localized by the AR platform, an important dimension is what \textit{content} will be shown at the identified locations~\cite{caetano_arfy_2022}. 
Previous work has focused on optimizing the presentation of digital content (e.g., existing app windows or widgets) on or around objects \cite{cheng_semanticadapt_2021, han_blendmr_2023, lindlbauer_context-aware_2019, lu_exploring_2022}, while others have investigated the use of tangible interaction with physical objects as a means to control such digital content \cite{zhu_mecharspace_2022, monteiro_teachable_2023, suzuki_realitysketch_2020}.
\textit{EditAR}~\cite{chidambaram_editar_2022} suggested capturing users and their interactions with objects nearby to create digital twins for later consumption in XR.
\textit{ProcessAR}~\cite{chidambaram_processar_2021} captures instruction demonstrations related to domain-specific objects by experts, so they can be viewed by novices in situ.

% but generalizable to a wider range of objects.

\textit{InfoLED}~\cite{yang_infoled_2019}, \textit{LightAnchors}~\cite{ahuja_lightanchors_2019}, \textit{BLEARVIS}~\cite{strecker_mr_2023}, and \textit{Reality Editor}~\cite{heun_reality_2013}  took this one step further to show truly \textit{object-specific content} in AR, however, this was specifically for electronic objects, such as their showing their battery level or device status in AR.
Researchers further suggested augmenting visualizations with dynamic AR content~\cite{chen_augmenting_2020, chulpongsatorn_augmented_2023, chulpongsatorn_holotouch_2023, xiao_imarker_2022}, however, such dynamic content was limited to printed documents or displays.

\vspace{0.2cm}
Our work builds upon these prior efforts by proposing a system that enables \textbf{object-specific content} and interactions for a much wider range of objects, regardless of their physical capabilities or pre-configured markers. We leverage advancements in spatial understanding via techniques like SLAM~\cite{davison_real-time_2003}, available in \textit{ARCore} and \textit{ARKit}, and machine learning models for object segmentation and classification to achieve this. This allows us to implement AR interactions with semantic depth, enabling contextually relevant information and actions for any object in the user's environment.

% The role of the UI (discoverability)\hl{ @Mar can add more explanation}
\begin{figure*}[t]
  \centering
  \includegraphics[width=.9\linewidth]{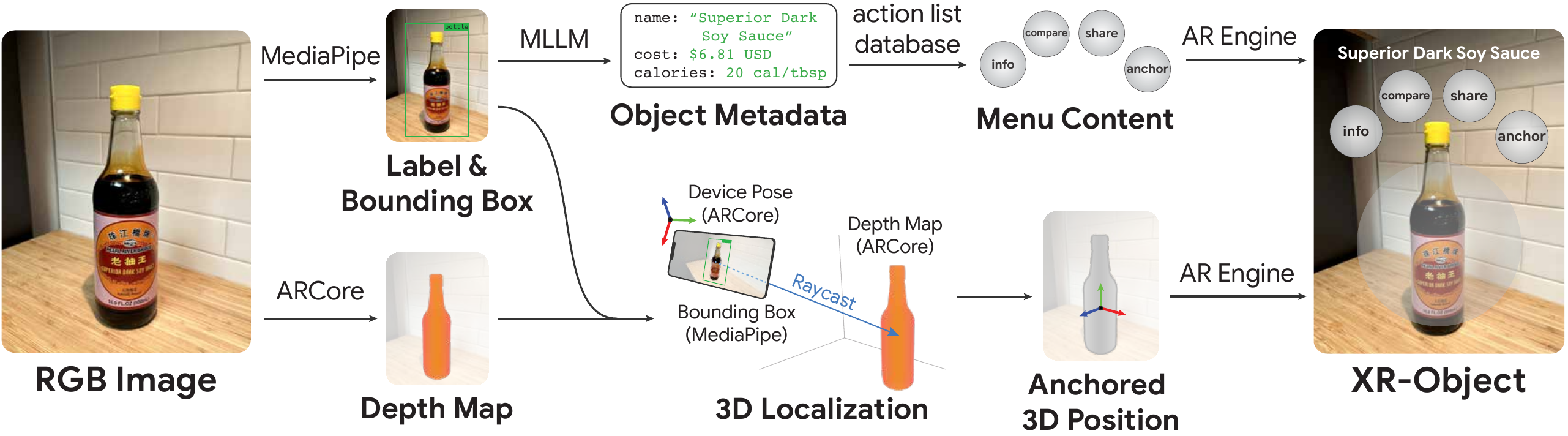}
  \caption{%\systemName~ processing pipeline: Live RGB images are processed through MediaPipe for object detection and through ARCore for spatial tracking and scene depth estimation. The bounding box of the detected object is then fed into MLLM for retrieving object metadata and also combined with the scene depth to compute the exact 3D anchor position of the object. The output of MLLM is used to populate the UI's menu content, which is then rendered by the AR engine at the 3D anchor position.
  The \systemName~ processing pipeline combines MediaPipe and ARCore for object detection and spatial tracking, respectively, integrates an MLLM for object-specific metadata retrieval and interaction, and renders UI content in 3D space.
  }
  \Description{The RGB image is processed for both MediaPipe labeling and ARCore depth map generation. The object metadata is retrieved from the MLLM. A raycast is used for 3D localization of the object. Finally, the menu content is anchored in 3D.}
  \label{fig:processing_pipeline}
\end{figure*}

\section{\systemName  ~Implementation}
\label{implementation}

\systemName~  leverages developments in spatial understanding via tools such as  SLAM, available in \textit{ARCore}~\cite{google_arcore_2024} and \textit{ARKit}~\cite{inc_arkit_2024}, and machine learning models for object segmentation and classification (\textit{COCO}~\cite{lin_microsoft_2015} via \textit{MediaPipe}~\cite{lugaresi_mediapipe_2019}), which enables us to implement AR interactions with semantic depth.
% have paved the way for more sophisticated interactions.
We also integrate a Multimodal Large Language Model (MLLM)
%, \textit{Google Gemini}~\cite{team2023gemini}, 
into our system, which further enhances our ability to automate the recognition of objects and their specific semantic information within XR spaces.

\paragraph{Platform}
Given the current constraints of AR headsets, particularly their limited developer access to real-time camera streams, we consciously targeted smartphones for our mobile prototype development. We aim to enable anyone to try out our open-source project on their phone. Modern smartphones use similar types of ARM-based mobile chipsets as AR headsets, yielding comparable performance for real-time computer vision tasks, while providing unrestricted access to their high-resolution cameras. This enables our application to identify objects in the user's environment and overlay digital information directly onto the physical world through the phone's display thanks to \textit{ARCore} and \textit{ARKit}.

\paragraph{Multimodal Interaction}
At the heart of \systemName~ is a multimodal large language model (MLLM)~\cite{yin_survey_2024} as well as a speech recognizer, which facilitate a rich interaction layer between the user and the objects. This model not only recognizes objects but also fetches and provides contextual information and actions relevant to the selected object. By integrating voice and visual inputs, our system offers a seamless and familiar interface for users to engage with their surroundings in novel ways.

\subsection{Design Considerations}

In developing the AOI paradigm for XR environments, we considered a range of design choices to enhance user interaction and system performance. These considerations are grounded in our review of related work and guided by our goal to seamlessly integrate digital functionalities with physical objects. Here, we explain our rationale behind key design decisions, contrasting them with alternative approaches and situating them within the broader discourse of HCI and AR.

\subsubsection{Object-Centric vs. App-Centric Interaction}

Traditional AR interactions often follow an app-centric model, where users must first open a specific application to access digital functionalities. In this model, users have to navigate through the app's interface to select categories or objects of interest, and in some cases, even upload pictures for analysis. Examples of app-centric interactions include standard \textit{ChatGPT}-style interfaces, where users input a query and an image, and \textit{Google Lens}, which requires users to open the app and manually select the objects they wish to interact with.

In contrast, our system prioritizes an object-centric approach, where interactions are directly anchored to objects within the user's environment. This means that users can immediately access digital functionalities by selecting an object, without the need to navigate through an app or input additional information. By leveraging advanced computer vision and spatial understanding techniques, our AOI framework enables users to seamlessly engage with the physical world as if it were a digital interface.
\new{We also note that, currently, while our research prototype does need to be installed as an application package, it aims for eventual native integration, similar to how QR scanning is now embedded in smartphones. On XR headsets, AOI could be enabled as an additional layer of the home space, in a similar manner as video passthrough can be toggled on and off.}

An object-centric approach would offer several advantages over app-centric models. Firstly, it provides a more natural interaction flow, as users can directly engage with objects in their surroundings without the cognitive burden of switching between the physical world and a digital app. Secondly, it minimizes the operational steps required to access digital functionalities, streamlining the user experience, enabling multi-tasking, and reducing friction. Finally, by anchoring interactions directly to objects, our AOI framework may offer a more immersive and seamless XR experience.

\subsubsection{World-Space UI vs. Screen-Space UI}
The choice between implementing a world-space UI versus a screen-space UI was informed by our aim to maintain spatial consistency and enhance user engagement with the XR environment. A screen-space UI, fixed relative to the user's viewpoint, could potentially obfuscate the immersive experience by detaching digital interactions from their physical context. Conversely, our adoption of a world-space UI, where digital elements are anchored to physical objects (akin to "billboards" in 3D graphics, i.e., user-facing 2D planes in a 3D space), ensures that interactions remain contextually grounded within the user's real-world environment.
We hope to minimize cognitive load by preserving spatial orientation and also leverage the natural human capability to navigate and interact with 3D spaces.

\subsubsection{Signaling Identified \systemName}
To mitigate visual clutter, a common issue in densely populated AR environments, we introduce the use of semi-transparent spheres, or "bubbles," as minimalist indicators of interactable objects. This design choice is based on the principle of minimalism and unobtrusiveness, ensuring that users are not overwhelmed by excessive digital information overlaying their physical surroundings. Bubbles serve as subtle prompts that an object is interactive to balance informational availability with spatial aesthetics.

% This enhances the user experience by maintaining a clear and uncluttered visual field, inviting users to engage with digital augmentations at their discretion.

\subsubsection{Fixed Number of Top-Level Categories and Actions}
The decision to implement a fixed number of top-level categories and actions within the system's UI was driven by considerations of usability and cognitive efficiency. Limiting the choice set helps mitigate decision fatigue and simplifies the interaction process, making it easier for users to navigate the system's functionalities. This design philosophy aligns with the Hick-Hyman Law~\cite{wu_hickhyman_2017}, which states that increasing the number of choices proportionally increases decision time. By streamlining the number of options available, we are also able to adopt a radial menu with constant reach distance \cite{samp_supporting_2010, pourmemar_visualizing_2019} instead of dropdown lists, and we facilitate quicker user decision-making and enhance the overall user experience.
% Furthermore, this approach aids in maintaining a clean and organized UI, crucial for avoiding overload and confusion in AR settings.
%\hl{<add why we use radial menus instead of dropdown, because more effective per Samp et al.~\cite{samp_supporting_2010} and Pourmemar et al.~\cite{pourmemar_visualizing_2019}>}

\vspace{0.2cm}

In summary, we aim to deliver a seamless, efficient, and immersive XR experience by opting for an object-centric interaction model, employing a world-space UI, utilizing visual bubbles for indicating interactability, and limiting the complexity of user choices.

\begin{figure*}[t]
  \centering
  \includegraphics[width=.85\linewidth]{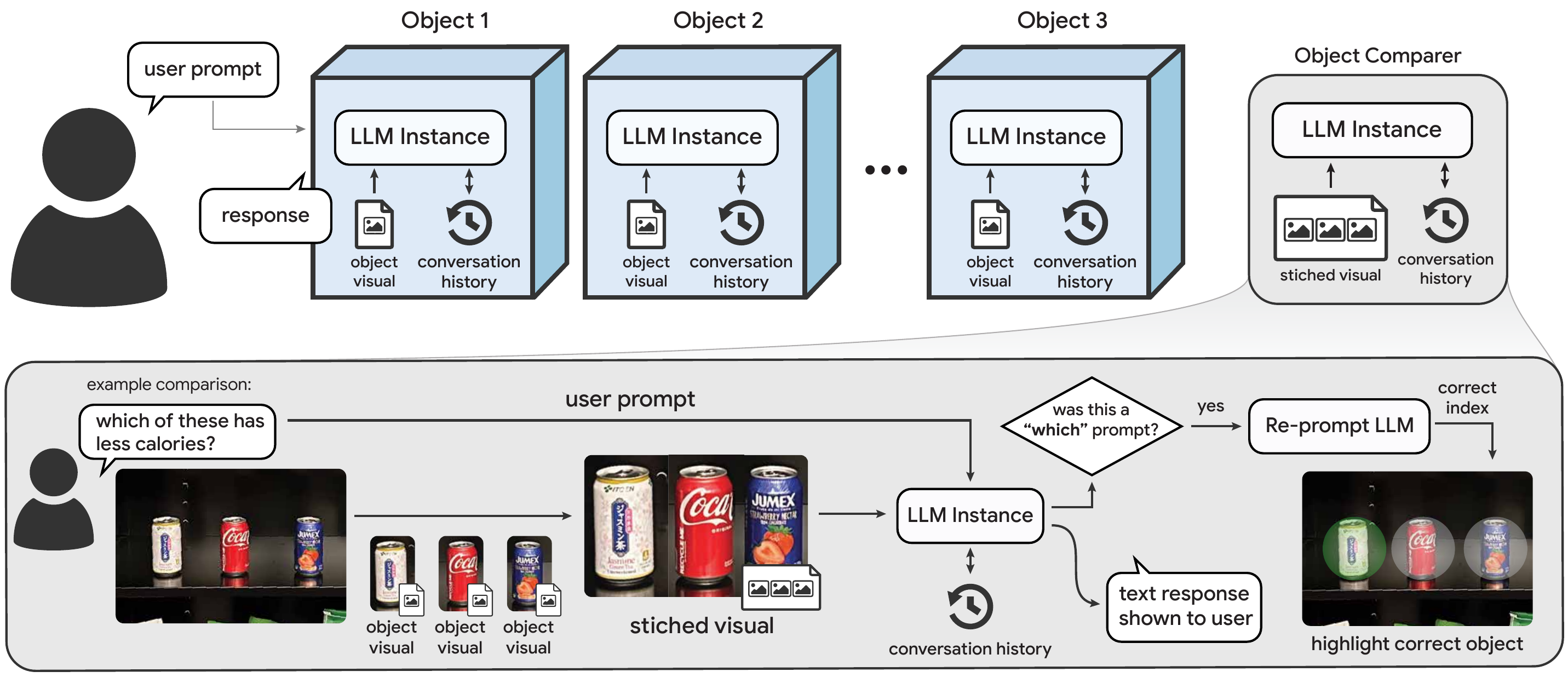}
  % https://docs.google.com/drawings/d/1xeq69ToK2dPpE7a94_TIjMK8BTyISSPlqq-qsIIndMg/edit?usp=drive_link
  \caption{ \systemName~ instantiates a dedicated MLLM instance for identified object in the scene. Object comparisons are executed by stitching together the relevant objects in the scene before passing the query to an MLLM instance. %\ruofei{see if you can come up with a better version.. for example, anything smarter in combining the conversation history? how to deal with long tokens? what if two objects are from completely different categories?}
  }
  %   \ruofei{can you replace all png icon with the bottle in figure 4?}
  \Description{Top: The user's prompt is supplied to the corresponding object. Each object, represented by a box in this diagram, has an LLM instance. There is also an object comparer running. Bottom: Inside the object comparer, the objects' images are stitched together to ask the LLM instance a question regarding multiple objects. The correct object is marked green in the real-world scene.}
  \label{fig:LLMObjectIntegrationFigure}
\end{figure*}

\subsection{Categories of Actions}
Our system facilitates fluid interactions with a single or multiple objects and enables users to take various digital actions, such as querying real-time information, asking questions, sharing the objects with contacts, or adding spatial notes.
Inspired by sub-menus in traditional context menus on desktop computing, we categorized our seven implemented actions into four categories, which we list below.

\begin{enumerate}[leftmargin=1cm]
    \item \textbf{Information}: provide an overview; ask a question
    \item \textbf{Compare}: ask to compare multiple objects within the view
    \item \textbf{Share}: send object to a contact; add to shopping list
    \item \textbf{Anchor}: notes; timer; countdown
\end{enumerate}

In the above list, the first two categories (\textbf{Information} and \textbf{Compare}) represent traditional \textit{Visual Question Answering (VQA)} tasks, while the other two (\textbf{Share} and \textbf{Anchor}) represent traditional \textit{widget} tasks.
We open-source our code on \textit{GitHub}\footnote{\systemName~ open-source project: \url{https://github.com/google/xr-objects}} and anticipate that the list of integrated actions will be extended in the future by the XR community.
%\textit{GitHub}\footnote{XR-Objects open-source project: \url{https://github.com/google/xr-objects}}

\subsection{System Architecture}
The implementation of \systemName~ involves a series of steps to augment real-world objects with functional context menus as illustrated in Figure ~\ref{fig:processing_pipeline}. These steps can be categorized as 
(1) detecting the objects, 
(2) localizing and anchoring onto the object, 
(3) coupling each object with an MLLM for metadata retrieval, and
(4) on user input, executing the actions and displaying the output.
We use \textit{Unity} and its \textit{AR Foundation}\footnote{Unity AR Foundation: \url{https://unity.com/unity/features/arfoundation}} to bring the necessary components for these steps together to build our system.
Below, we detail the components and processes that constitute this framework.

\subsubsection{Object Detection and Classification}
The foundation of \systemName~ is its robust object detection module, which leverages the capabilities of the \textit{Google MediaPipe} library~\cite{lugaresi_mediapipe_2019}. This module employs a convolutional neural network (CNN) optimized for mobile devices, providing \textit{on-device} and real-time classification of objects within the user's camera feed. The system detects objects by providing a \textit{class label} (e.g., “bottle”, “monitor”, “plant”) and generating \textit{2D bounding boxes}, which serve as preliminary spatial anchors for subsequent AR content.
The current CNN model is based on the \textit{COCO} dataset~\cite{lin_microsoft_2015} which provides 80 labels.

To prioritize user privacy and data efficiency, \systemName~ further processes only those object regions that are of relevance to the user's current interaction context.
% Only if the returned object label is of interest to the user, then we crop and only send that region of the user’s AR view to the server to get further metadata of the object, as explained in the next section.
% This approach also preserves the privacy of users:
% We do not send the whole user view or every frame to the LLM-based visual query system on the cloud.
% Furthermore, for the privacy of surrounding users,
For instance, even though \textit{MediaPipe} inherently also identifies people in the scene, a region classified as a “person” by the on-device model is not sent to the MLLM-based cloud query system to preserve the privacy of users in the surroundings.
Similarly, other classes of objects that are relevant or irrelevant to the user can be customized %by the user themselves or
depending on the AR application (e.g., a plant species search AR app could only run queries for the “plant” class).

\subsubsection{3D Localization and Anchoring}
With the object identified, \systemName~ proceeds to generate AR menus. These menus are spatially anchored to the objects using a combination of the initial 2D bounding boxes and depth information of the scene.
We us raycasting to translate the 2D object locations into precise 3D coordinates.

In our system, we used the depth map on the phone~\cite{du_depthlab_2020} generated through depth from motion~\cite{valentin_depth_2018} by \textit{ARCore}. Because the location returned by the object detector from the previous step is in 2D screen space, we raycast from this point toward the depth map to “hit” the object and find the corresponding 3D object location in world space, as shown in Figure~\ref{fig:processing_pipeline}.

At the computed 3D location, we instantiate our object proxy template, which was developed as a \textit{prefab} (i.e., fully configured \textit{GameObjects} saved in the AR project for reuse) in \textit{Unity}.
The AR engine ensures that the object proxy stays anchored even when the user changes their view angle.

The object proxy contains the object's context menu, however, before the user selects the object, it shows up as only a small semi-transparent sphere on the object, which signals to the user that the object has been recognized by the system. Only when the user taps this sphere, the full context menu is shown, otherwise, the menu remains hidden to avoid visual overload.

%Our algorithm also includes extra steps to ensure the object proxies do not get spawned at undesired locations or get erroneously duplicated for the same object.

Our algorithm includes additional steps to prevent object proxies from being spawned in unintended locations or duplicated erro- neously for the same object.

\subsubsection{Coupling Each Object with a MLLM}
We couple a MLLM with each identified object; thus, we run one LLM conversation instance per object, as shown in Figure~\ref{fig:LLMObjectIntegrationFigure}. 
We use the cropped bounding box from the first step as the visual input to the MLLM.
We also store the conversation history by referring to a conversation ID internally.
This object-specific approach enables the MLLM to provide detailed information about the object that extends far beyond the capabilities of traditional object classifiers. For example, it can furnish the object with a wide array of data, including but not limited to product specifications, historical context, and user reviews. As demonstrated in Figure~\ref{fig:processing_pipeline}, the system is capable of recognizing an object as "Superior Dark Soy Sauce," rather than merely identifying it as a "bottle"—the generic label typically assigned by standard object detection processes in the preceding step.

For our MLLM, we use \textit{PaLI}~\cite{chen_pali_2023}, which runs in the cloud and takes as input the captured region of interest (i.e., object’s cropped image). The MLLM system is able to simultaneously query the Internet (\textit{via Google Search}) to retrieve additional metadata about the object, e.g., prices and user ratings in the case of a product.

\subsubsection{Menu Interaction}
Interactions within \systemName~ are facilitated through a multimodal interface, supporting both touch and voice inputs. This flexible interaction model allows users to engage with the system in a manner that best suits their preferences and the current context. For voice interactions, the system incorporates a speech recognition engine\footnote{Speech Recognizer: \url{https://github.com/EricBatlle/UnityAndroidSpeechRecognizer}}, which enables the processing of natural language commands and queries.
As the feedback mechanism, certain user actions, such as selecting a menu option or asking a question, are reflected in the panel overlaid on the object.

When the object is selected by the user, the actions, described in the previous section, are shown.
Once an action's button is tapped, the interaction starts in a panel overlaid on the object.

\paragraph{Information retrieval}
For the actions that retrieve real-time data (e.g., getting the answer to the user's question), we use the object's MLLM instance.
For instance, when the "info" button is selected, the MLLM-returned object summary is shown. We use the following prompt to create an object summary:

\begin{displayquote}
\small \textit{Provide the information from the following list that makes sense for this object. Fill in the missing ``…" using info from the Internet. Exclude the one that are irrelevant. Divide the relevant ones with a ``*". * Price: … (give price+vendor+score/ rating) * Cheaper alternatives: name - price * Main ingredients: … (top 2) * Calories: … * Allergens: … * Instructions: … (short) * Care: …(if fashion/tool/plant). Use extremely short answers and exclude answers that are `None' or `n/a' or `irrelevant'. Limit to 30 words.}
\end{displayquote}

\vspace{0.2cm}

If the user wants to ask a more specific question, they can tap the “ask a question" button, and directly speak out a question.
% Such actions which are interacted with using speech use \textit{Android}'s native speech recognition service 
% as a \textit{Unity} library~\footnote{\url{https://github.com/EricBatlle/UnityAndroidSpeechRecognizer}}.

\paragraph{Object comparisons}
For the object \textit{compare} functionality, we use a dedicated  “object comparer” method, which allows us to compile multiple identified objects’ information and provide the combined result as input to a dedicated MLLM instance. As shown in Figure~\ref{fig:LLMObjectIntegrationFigure}, the object comparer stitches all objects’ images together and provides its MLLM instance when the user asks a question $prompt_{user}$ about multiple objects using the “compare" button. The returned response is shown to the user.

If the user’s prompt is a “which” question, the object comparer also executes a follow-up MLLM query under the hood to help visualize the results for this “filtering” type of user question. For this reprompting, we augment the user’s prompt $prompt_{user}$ with a sub-prompt $prompt_{indexing}$ such as:

\begin{displayquote}
\small 
\textit{Considering that the items are ordered from left to right with the first object being index 0, tell me ONLY the correct indices, written as numbers.}
\end{displayquote}

\vspace{0.2cm}

Thus, the MLLM returns only the right indices, which we use to mark the relevant objects in the AR view as shown in the bottom-right screenshot in Figure \ref{fig:LLMObjectIntegrationFigure}.

\section{Preliminary Evaluation}
\label{evaluation}
 %We conducted a user study to understand the design space for object-centric AR context menus and widgets. We found that users found context menus and widgets to be useful for a variety of tasks, such as getting information about objects, interacting with them, and personalizing them. We also found that users had a preference for context menus that were lightweight and presented a limited number of options, and for widgets that were visually appealing and easy to interact with.

%This paper presents a study of the design space for object-centric AR experiences with a focus on two key interaction mechanisms: context menus and widgets. Context menus provide users with a way to access a range of actions that are relevant to a particular object. Widgets are small, interactive elements that can be placed on or around objects to provide users with related information or functionality.

We conducted a user study comparing \systemName{} to a state-of-the-art MLLM assistant interface (Gemini app\footnote{Google Gemini: \url{https://play.google.com/store/apps/details?id=com.google.android.apps.bard
}}), referred to as "Chatbot" from here on, for contextual information retrieval and object-centric interaction. Participants were asked to perform a number of timed VQA tasks and widget interactions in simulated grocery shopping and at-home scenarios, and provided feedback on their experience with each interface through a survey.

% Multimodal-LLM powered by chatbot (Condition Chatbot)

% participants
\subsection{Methods}

\vspace{0.2cm} \noindent
\textit{\textbf{Participants}}.
We recruited 8 participants (6 male, 1 female, 1 preferred not to disclose) between the ages of 25-45. All were fluent English speakers (4 native), were regular shoppers, and all but 1 had used smartphone-based AR at least once.

% scenario & tasks

% \subsection{Scenario \& Tasks}

\begin{figure}[h]
  \centering
  \includegraphics[width=\linewidth]{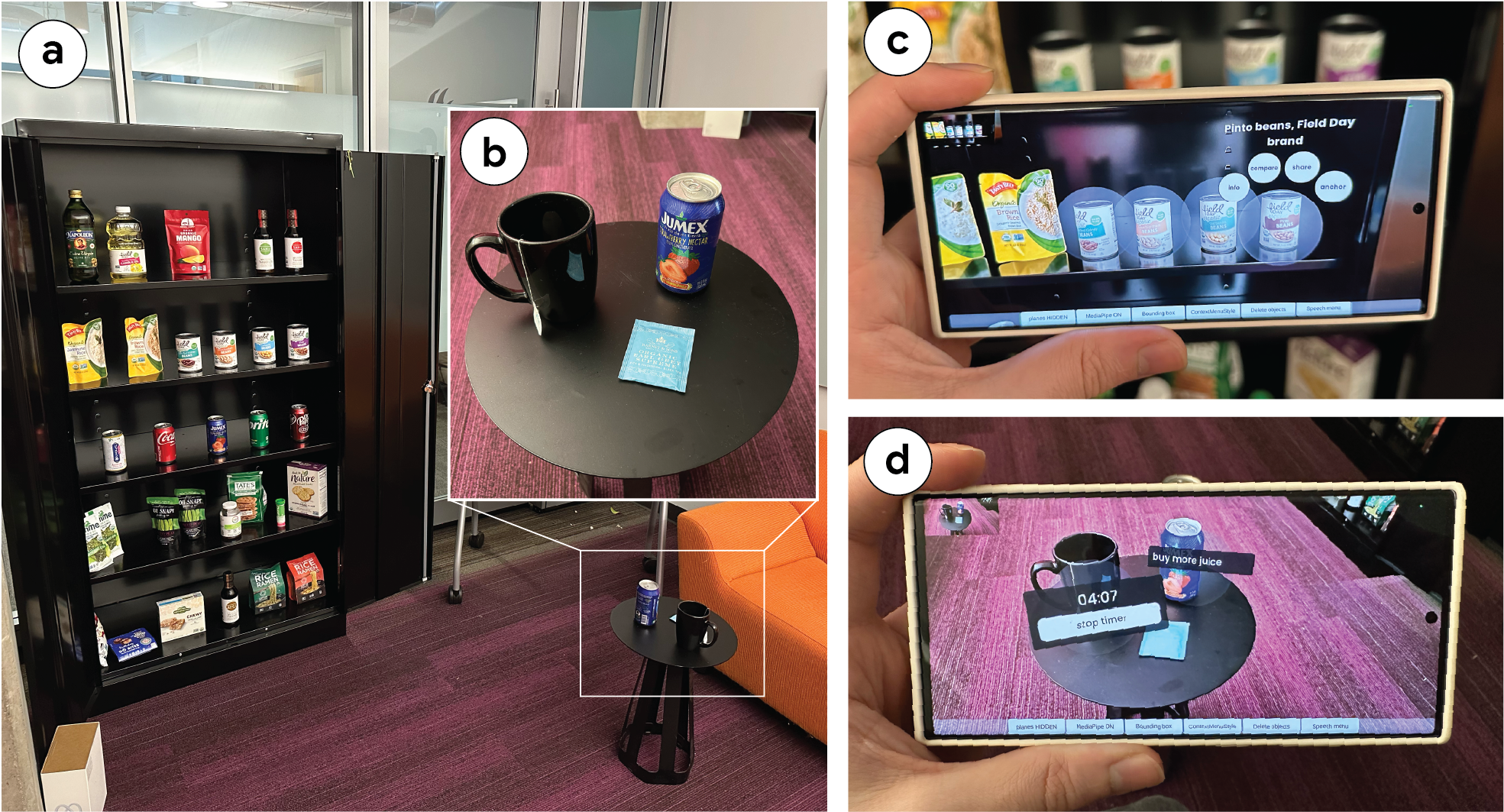}
  \caption{User study setup with mock grocery store (a) and at-home (b) environments. Examples of using \systemName~ in each case are shown in (c) and (d), respectively.}
  \Description{(a) Shelves wıth multiple products. (b) A small table with a mug and a can. (c) A phone shows the objects on the shelf in the UI. (d) The same UI looking at the objects on the table.}
  \label{fig:study-setup}
\end{figure}

\vspace{0.2cm} \noindent
\textit{\textbf{Task \& Procedure}}.
%We designed a scenario consisting of a simulated grocery shopping experience (\autoref{fig:study-setup}a) followed by an at-home experience (\autoref{fig:study-setup}b) in which users complete a set of 6 tasks using either \systemName~ or Chatbot. The task categories included: (\textbf{T1}) 2-object Search, (\textbf{T2}) N-object Search, (\textbf{T3}) Share Object, (\textbf{T4}) Get Info (Single Object), (\textbf{T5}) Set/Anchor Timer, and (\textbf{T6}) Create/Anchor Note. We created two variants of this scenario (\textbf{A} and \textbf{B}), each with the same categories but slightly varied tasks. The full set of tasks is presented in \autoref{tab:study-tasks}. 
We designed a scenario consisting of a simulated grocery shopping experience (\autoref{fig:study-setup}a) followed by an at-home experience (\autoref{fig:study-setup}b) in which users complete a set of 6 tasks using either \systemName~ or Chatbot. \new{The tasks (\textbf{T1}-\textbf{T6}) are listed in \autoref{tab:study-tasks}, and included retrieving/comparing information about multiple objects, sharing objects, and anchoring widgets. % to objects.
}

\begin{table}[h]
    \centering
    \includegraphics[width=1\linewidth]{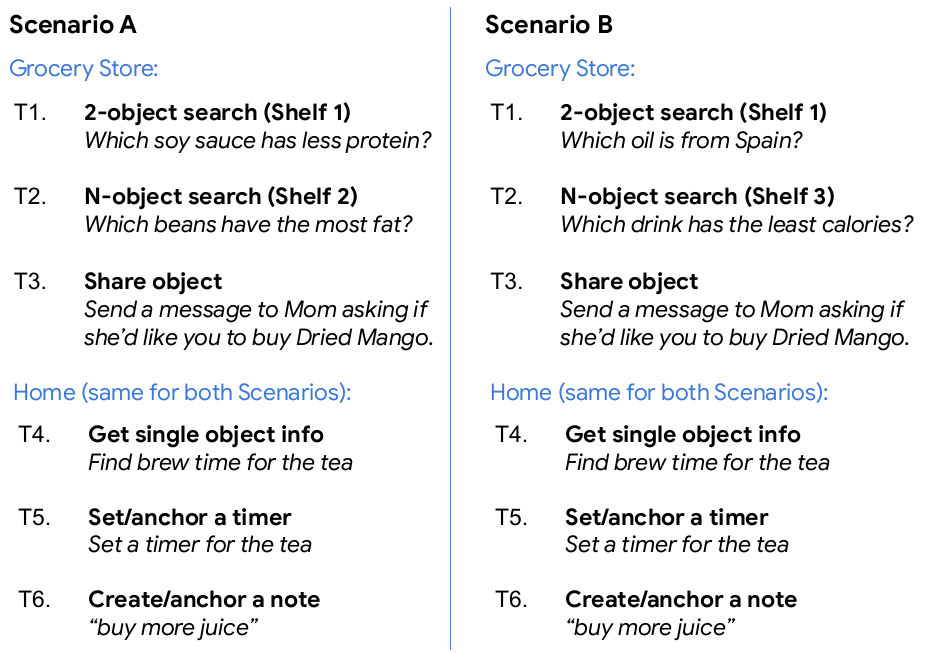}
    % https://docs.google.com/drawings/d/1uwDrSxORCozizsxSsri3T4MCQCvonQB1Iya6EIjNLWE/edit
    \caption{Task descriptions given in the user study.}
    \label{tab:study-tasks}
\end{table}

For \systemName, participants were instructed to use the \textbf{Compare} feature for \textbf{T1} and \textbf{T2}, the \textbf{Share} feature for \textbf{T3}, the \textbf{Info} feature for \textbf{T4} and the \textbf{Anchor} feature for \textbf{T5} and \textbf{T6}. For Chatbot, participants were instructed to take/upload a photo along with a query (using their preferred method of text \new{and/or} voice) to the chat for \textbf{T1}, \textbf{T2}, and \textbf{T4}. For the remaining tasks, participants were told to use Chatbot as if it were connected to a smartphone assistant.

% procedure
% \subsection{Procedure}

First, participants were first given a brief introduction to the study, provided informed consent, and filled out a demographics survey. The experimenter then walked through the functionality of both \systemName~ and Chatbot, and participants then completed a set of sample tasks on an object not included in the study.

Each participant completed the tasks in both scenarios \textbf{A} and \textbf{B}. The ordering of both condition (\systemName, ~Chatbot) and scenario (\textbf{A},~\textbf{B}) were counterbalanced between participants to prevent ordering effects. Once the tasks were completed, participants were free to use the tool freely for up to five additional minutes however they chose. Participants then completed a post-condition survey on their qualitative experience. %(Appendix \ref{post-condition survey}).

After a two minute break, this was repeated with the remaining condition and scenario. Upon completing the tasks with both conditions, participants completed a final survey comparing interactions with \systemName~ and Chatbot (Appendix \ref{post-study survey}). Overall, the study took approximately 45 minutes.

% \subsection{Measures}
\vspace{0.2cm} \noindent
\textit{\textbf{Measures}}.
% \subsubsection{Task Completion Time}
As one of our measures, we recorded the \textbf{time required to complete tasks} \textbf{T1}-\textbf{T6} for each condition as a measure of overall system performance.

% \subsubsection{Post-Condition HALIE Survey}
Following the completion of all tasks in a given condition (\systemName, Chatbot), participants completed a survey evaluating their interactions, adapted from the \textbf{Human-AI Language-based Interaction Evaluation (HALIE)} framework proposed by Lee et al. \cite{lee_evaluating_2022}. Participants rated their agreement with the following statements (among others) on a 5-point \textit{Likert} scale. Due to space constraints, the complete survey is provided in Appendix~\ref{post-condition survey}.

% \subsubsection{Post-Study Form Factor Survey} \label{post-study survey}

While the user study was conducted on a smartphone due to limitations of head-mounted display (HMD) camera access, our vision for \systemName~ is for it to run entirely on the HMD. Therefore, we conducted a \textbf{post-study form factor survey} in which participants envisioned interacting with \systemName~ on an HMD (e.g., \textit{Apple Vision Pro}). The survey questions (provided in Appendix~\ref{follow-up survey}) were based on HALIE, but formulated as a forced-choice comparison between the Chatbot and \systemName~ interaction paradigms.

% \subsection{Analysis}
\vspace{0.2cm} \noindent
\textit{\textbf{Analysis}}.
% \subsubsection{Task Completion Time}
We analyze \textbf{completion time} using traditional \textit{t-tests}, and confirm normality via \textit{Shapiro-Wilk Test}.
%The completion time for tasks, a continuous variable, was analyzed using traditional t-test comparisons, under the assumption of normal distribution. The normality of our dataset was verified through the \textit{Shapiro-Wilk} test.
%Time is a continuous measure that can be analyzed with traditional \textit{t-test} comparisons, if it is normally distributed. We confirmed normality of our data using \textit{Shapiro-Wilk Test}.
% \subsubsection{Distribution Skewness}
\textbf{Skewness} ($\gamma_{1}$) quantifies the asymmetry of a given distribution's shape. For a normal distribution, values of $|\gamma_{1}|$ < 0.5 indicate an approximately symmetric distribution. Values of 0.5 < $|\gamma_{1}|$ < 1  suggest moderate skewness, while $|\gamma_{1}|$ > 1 suggests a highly skewed distribution. This statistical approach, in contrast to visual methods like histograms, is particularly useful for analyzing distributions in \textit{Likert}-scale questionnaires \cite{king_statistical_2018}.

% \subsubsection{Generalized Linear Model}
To analyze the data derived from our forced-choice questionnaires, we use a \textbf{Generalized Linear Model (GLM)} based on the Logit Binomial distribution. Unlike regular linear models, GLMs enable regression beyond Gaussian distributions. Considering this data follows a Bernoulli distribution (i.e., datapoints are 0 or 1), our GLM is effectively a log-odds model. %I.e in our GLM each outcome of Y (dependent variable) is assumed to be generated by our particular probability distribution. 

\subsection{Results}

\subsubsection{Time}
On average, participants using \systemName~ required significantly less time (M=217.5s, SD=58s) to complete all tasks when compared to the participants using Chatbot (M=286.3s, SD=71s), illustrated in Figure \ref{fig:time} and confirmed by a paired t-test (t=-2.8, df=5, p=0.01). This translates to a roughly 31\% in task completion time on average for Chatbot users compared to \systemName.

\subsubsection{HALIE Survey}
\begin{figure}[h]
  \centering
  \includegraphics[width=0.8\linewidth]{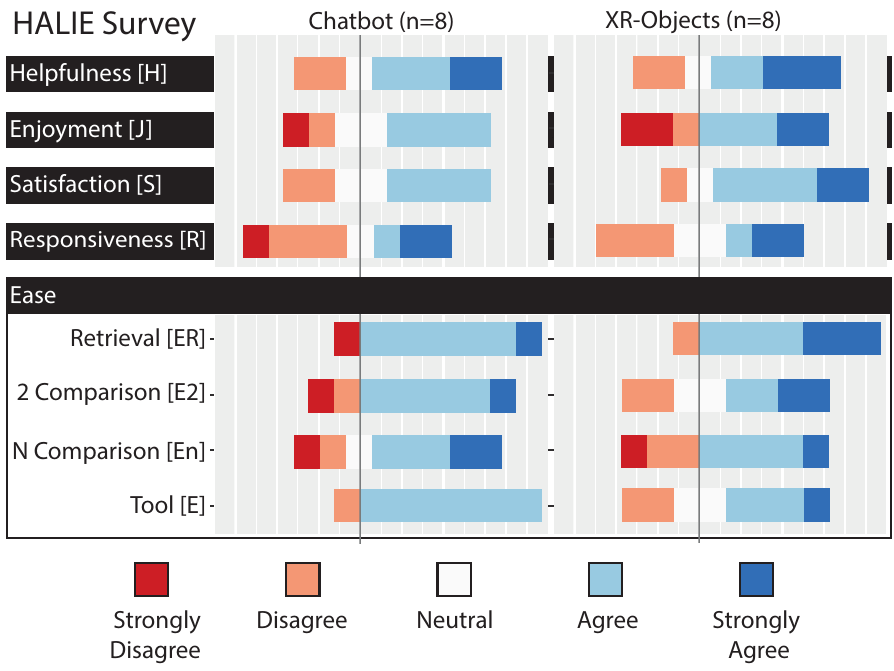}
  \caption{\textit{Likert}-scale results of the HALIE survey.}
  \Description{The HALIE survey rated helpfulness, enjoyment, satisfaction, and responsiveness. For ease, 4 conditions were rated: retrieval, 2 comparison, N comparison, and tool.}
  \label{fig:halie}
\end{figure}

We analyze the HALIE survey results using both traditional non-parametric tests for ordinal data and skewness calculations to assess the distribution of responses (Figure~\ref{fig:halie}).

% We study the survey results using both traditional non-parametric testing for non-continuous data, as well as sample distribution statistics using skewness on the results of our study survey (Figure \ref{fig:halie}). 

 While we find no significant differences between Chatbot and \systemName~on any factor of the HAILE survey (\textit{Wilcoxon Paired tests}), we find that both approaches of MLLM-enabled real-world search (either via Chatbot or \systemName) appear positively rated. Thus, we proceed with a skewness analysis. The most significant skewness of the questionnaire data were found on the questions regarding Ease of Use, and Satisfaction. In particular, Ease of Information Retrieval showed both conditions were highly skewed: \systemName~ ($\gamma_{1}=1.19$) Chatbot ( $\gamma_{1}=1.8$), making a strong case for MLLM-enabled information retrieval in any form.

 Further exploration shows that the responses for Tool Ease were highly skewed for Chatbot ($\gamma_{1}=2.25$). However, those same skews weren't sustained on the \systemName~ condition ($\gamma_{1}=0.03$). We hypothesize that this is because \systemName~ is a research prototype, while the Chatbot used was a fully released product. Nevertheless, we found a moderate skewness on the questionnaire results for the Satisfaction metric only for our prototype \systemName~ ($\gamma_{1}=0.7$), but not for the Chatbot condition ($\gamma_{1}=0.4$).

% Overall these results seem to make a strong case for embedded phone apps that enable multimodal search in either XR-Objects or Multimodal LLM format.

\subsubsection{Form Factor (Phone or HMD)}

The results of the form factor survey are summarized in Figure \ref{fig:formfactor}. We applied a GLM with two factors: \textit{FormFactor} (phone, HMD) and \textit{Question} (across 12 questionnaire levels), assuming a binomial distribution. Given that the dependent variable's responses were binary (\systemName~ or Chatbot), traditional linear models were inadequate as they are tailored to fit Gaussian distributions only. The GLM approach allowed for fitting a Bernoulli distribution and conducting appropriate tests. Our analysis revealed a significant \textit{FormFactor} effect $ F(191, 179) = 1.917, p<7.05e-08, \eta^2 = 3.8$.

To further assess the model's effectiveness in predicting Form Factor, we examined the model deviance \((-2LL = 209)\) and compared it against the null model's deviance, which assumed Form Factor was not a consideration \((-2LL = 253)\). This comparison demonstrated that our model was more adept at accounting for the variance, where a higher deviance signifies a poorer fit. A chi-square test contrasting the two models yielded a significant difference, with \(\chi^{2} = 1.6 \times 10^{-5}, df = 12\).

  These findings show a clear preference for \systemName~ in the context of the HMD form factor. Conversely, when using a phone, participants' preferences between the AI tools (Chatbot or \systemName) were split, validating our hypothesis regarding the optimal form factor for tools like \systemName.

\subsection{Qualitative Feedback}
\new{We provide key insights from the responses that participants provided in their completed surveys.}

\vspace{0.2cm} \noindent
\new{\textit{\textbf{Helpfulness \& Efficiency}}.
Users valued the system's streamlined interactions:
\textit{"It saves me from looking up info myself... I just ask and it finds the info for me"} (P1).
\textit{"Object selection and comparison was very intuitive; it was easy to get information in exactly the context I needed"} (P3).}

\vspace{0.2cm} \noindent
\new{\textit{\textbf{Comparative Advantage}}.
The direct comparison with existing solutions like \textit{Google Lens} and traditional LLMs was enlightening: 
\textit{"I was able to complete the same tasks much faster + easier"} (P1).
\textit{"This has a lot more options, and is more flexible [in] what information it can provide"} (P2).
\textit{"Comparing products is very helpful... less wordy than [Chatbot] and gives an answer"} (P2).}

\vspace{0.2cm} \noindent
\new{\textit{\textbf{Possibilities for UX Improvement}}.
Several participants pointed out ergonomic challenges, e.g., the need to hold the phone at eye level, which informs future glass interactions as discussed in Section~\ref{discussion}:
\textit{"have to raise the phone at eye level, which is tiring"} (P1).}

\section{Applications}

\begin{figure}[h]
  \centering
  \includegraphics[width=0.83\columnwidth]{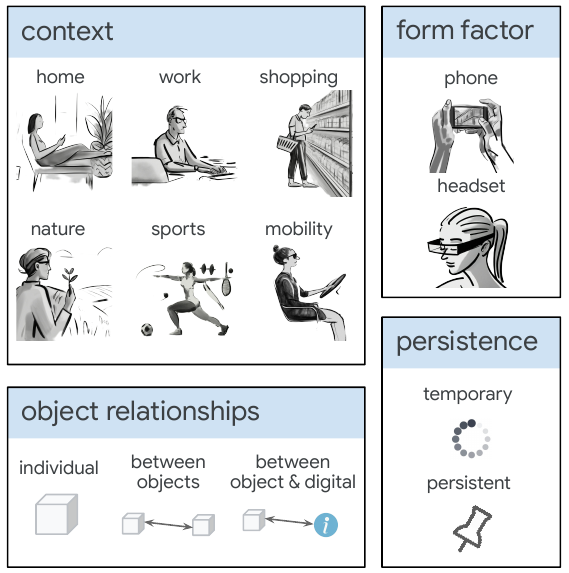}
  \caption{Envisioned design space for \systemName~ use cases. % \ruofei{I have never heard application space in this way. Any reference? partly from Ad hoc UI full paper but the form factor are limited, and context seem trivial with sketches, I suggest to remove.} 
  }

  \Description{Context: home, work, shopping, nature, sports, mobility. Object relationships: individual, between objects, between object & digital. Persistence: temporary vs. persistent.}
  \label{fig:applications}
\end{figure}

Through AOI, we envision \systemName~ to be useful across a variety of real-world applications. By enabling in situ digital interactions with non-instrumented analog objects, we can expand their utility (e.g., enabling a pot to double as a cooking timer), better synthesize their relevant information (e.g., comparing nutritional value), and overall enable richer interactivity and flexibility in everyday interactions. Next, we present five example application scenarios from a broad design space we envision as illustrated in Figure~\ref{fig:applications}, highlighting the value of \systemName~ and its potential use cases.

\subsection{Cooking}

Unlike traditional cooking aids, which rely on static recipes or digital screens detached from the cooking environment, \systemName~ integrates digital intelligence directly into the kitchen, making the cooking process informative and engaging (Figure \ref{fig:teaser}). In our augmented cooking app, as the user places ingredients on the kitchen counter, our system recognizes each item and projects relevant information, such as nutritional facts, potential allergens, or freshness indicators, directly on the ingredients. Users can interact through voice commands or touch elements to ask about potential recipes or to compare ingredients. Using stopwatch or countdown timers, the system embeds the guidance into the cooking space itself. The user can further share the final product with their contacts.

\new{This scenario shows how \systemName~ can assist users with multi-step tasks. 
For instance, P5 noted: “\systemName\textit{'s main benefit compared to Lens is spatializing output, [which] is super helpful for labeling multi-step tasks like cooking or mechanical fixing}."
While we illustrated cooking in this example, we envision our system can scale to further multi-step use cases, such as mechanical fixing.}

\begin{figure}[b]
  \centering
  \includegraphics[width=1\linewidth]{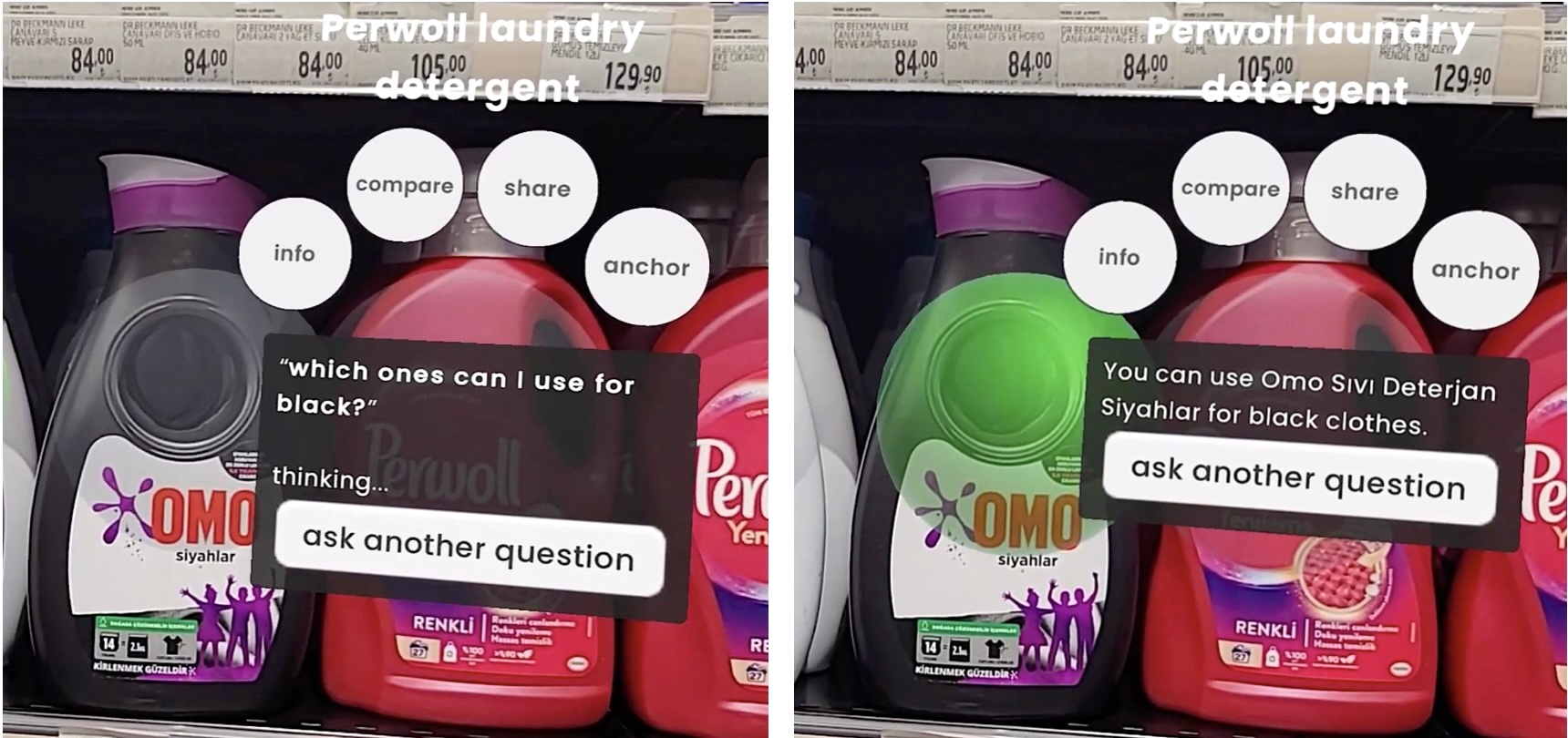}
  \caption{Real-time spatial assistance in selecting the appropriate laundry detergent at a store.}
  \Description{Left: detergents on a shelf. Right}
  \label{fig:shopping}
\end{figure}

\subsection{Shopping}
\new{As discussed in Section~\ref{evaluation}, \systemName~ can serve as an assistant when browsing and comparing items.
For instance, at a store, a user might want to get further information about a product, such as the unit price, calorie information, or cheaper alternatives. Figure~\ref{fig:shopping} shows an example where a tourist asks which of the detergents is suitable for black clothes, as the product information is written in a foreign language.
In the future, these experiences could be further personalized by, e.g., automatically filtering all the products identified in the scene  based on the user's personal profile and recommending the one that best suits their needs~\cite{xu_arshopping_2022}.}

\subsection{Discovery}

\begin{figure}[h]
  \centering
  \includegraphics[width=.95\linewidth]{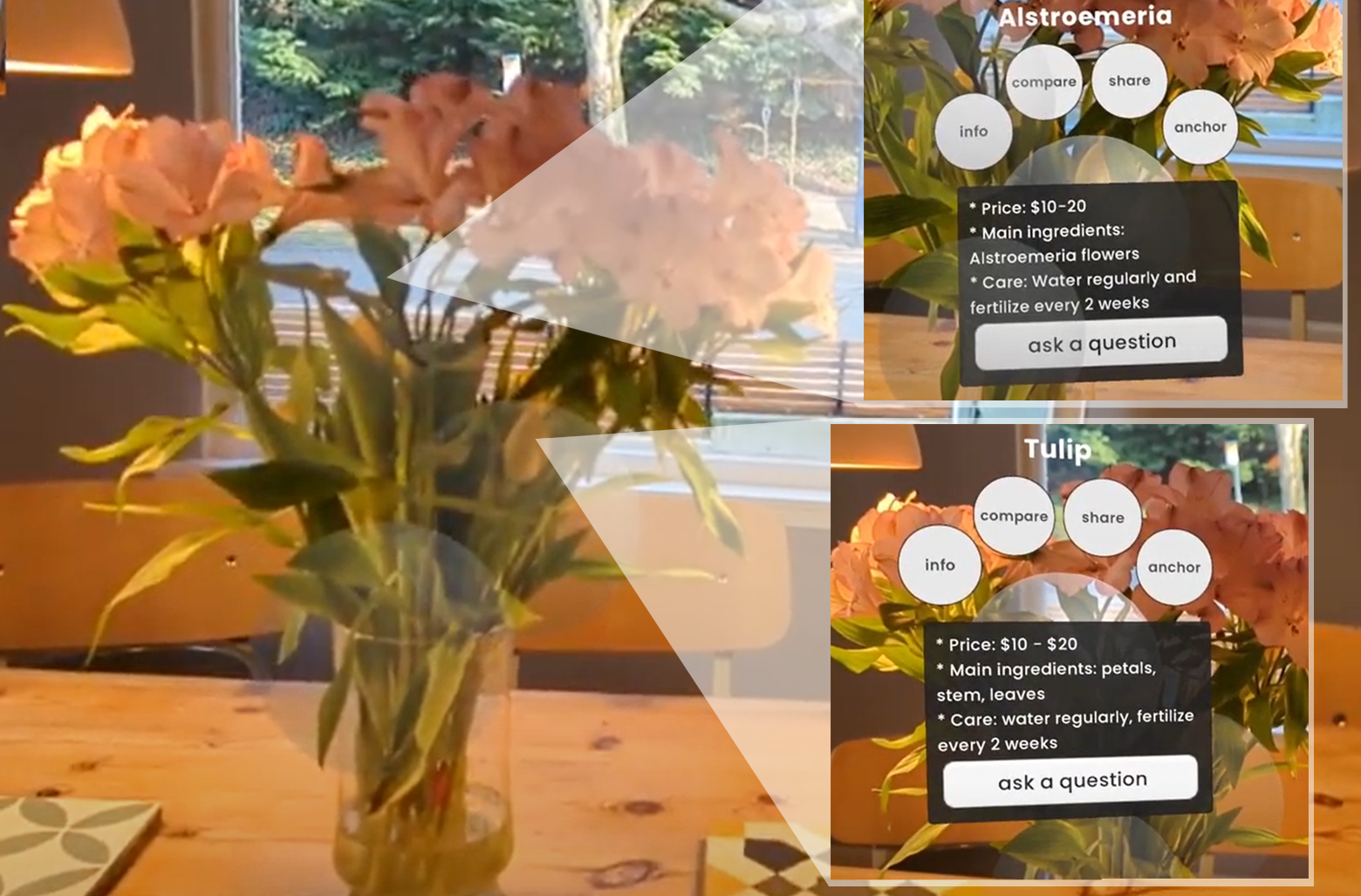}
  \caption{Discovering plant species in the environment through spatial, on-demand MLLM queries.}
  \Description{A bundle of flowers in a vase. The user can look up different species such as tulip vs. alstroemeria.}
  \label{fig:discover}
\end{figure}

\systemName~ enables users to discover new information about their surroundings by simply pointing their device at objects of interest. As shown in Figure \ref{fig:discover}, a user points their device at a vase containing different flowers and instantly receives information about each flower type, including its name, average price, or care instructions. This on-demand, spatial discovery could transform everyday objects that typically go unnoticed into avenues for appreciation. %understanding, learning, and appreciation. 

% By leveraging the power of object-centric MLLMs, \systemName~ can provide users with relevant facts, historical context, and even creative insights related to the objects they encounter.

\subsection{Productivity}

\begin{figure}[b]
  \centering
  \includegraphics[width=.95\linewidth]{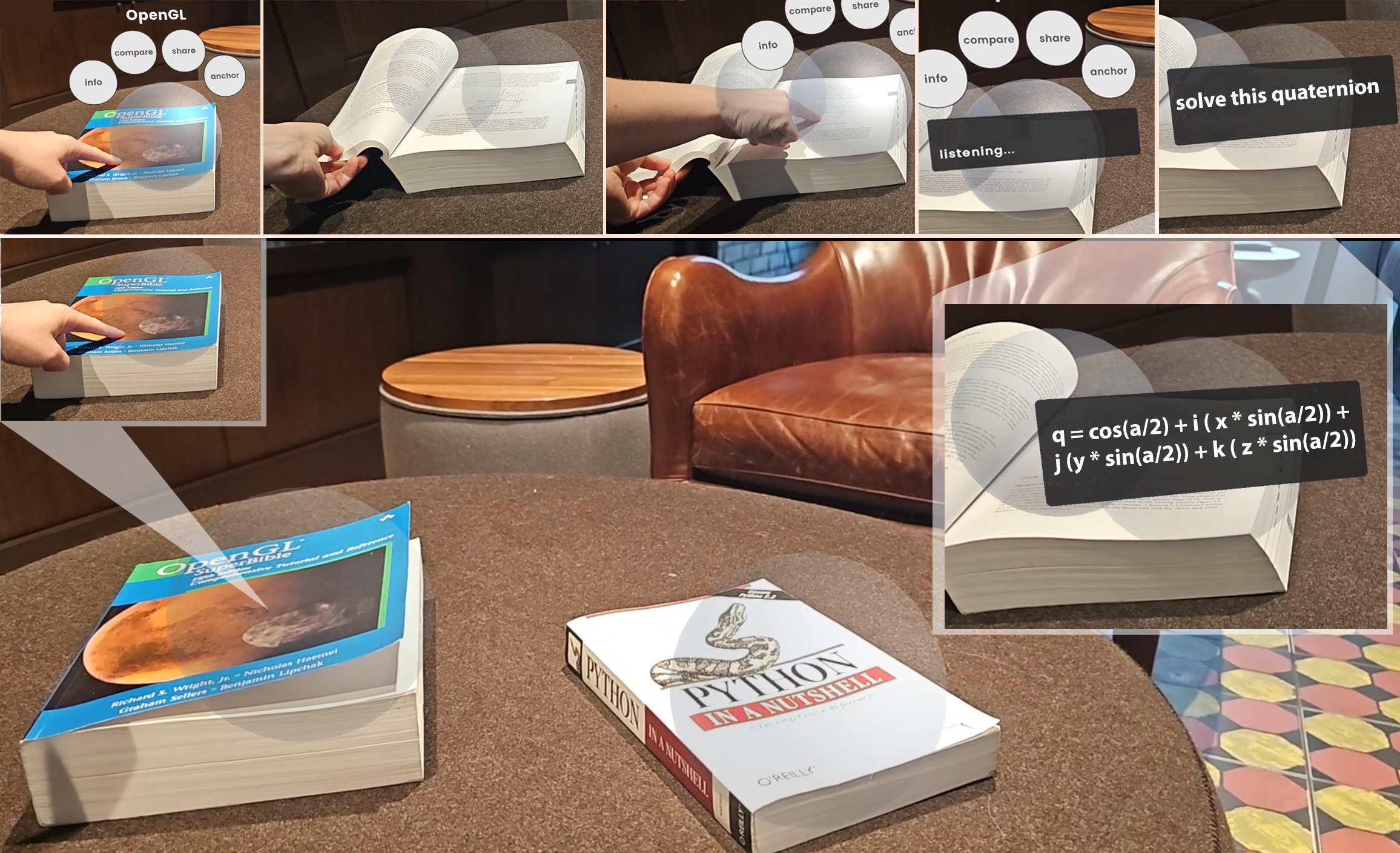}
  \caption{Interaction mock-up of using \systemName~ for productivity on an HMD with direct touch interactions enabled.}
  \Description{Two books on a table. The book on the left is about OpenGL. The user points at a page to ask a question.}
  \label{fig:productivity}
\end{figure}

For productivity, we envision \systemName~ could enhance physical documents with digital capabilities such as information retrieval and content anchoring. In Figure~\ref{fig:productivity}, a user reading a text book asks how it can be used to solve a particular type of equation, and anchors the response to the textbook for future reference. With further capabilities such as real-time optical character recognition (OCR) to digitize text, users could store and share digital copies of their physical documents for versioning and collaboration.

\subsection{Learning}

\begin{figure}[h]
  \centering
  \includegraphics[width=.95\linewidth]{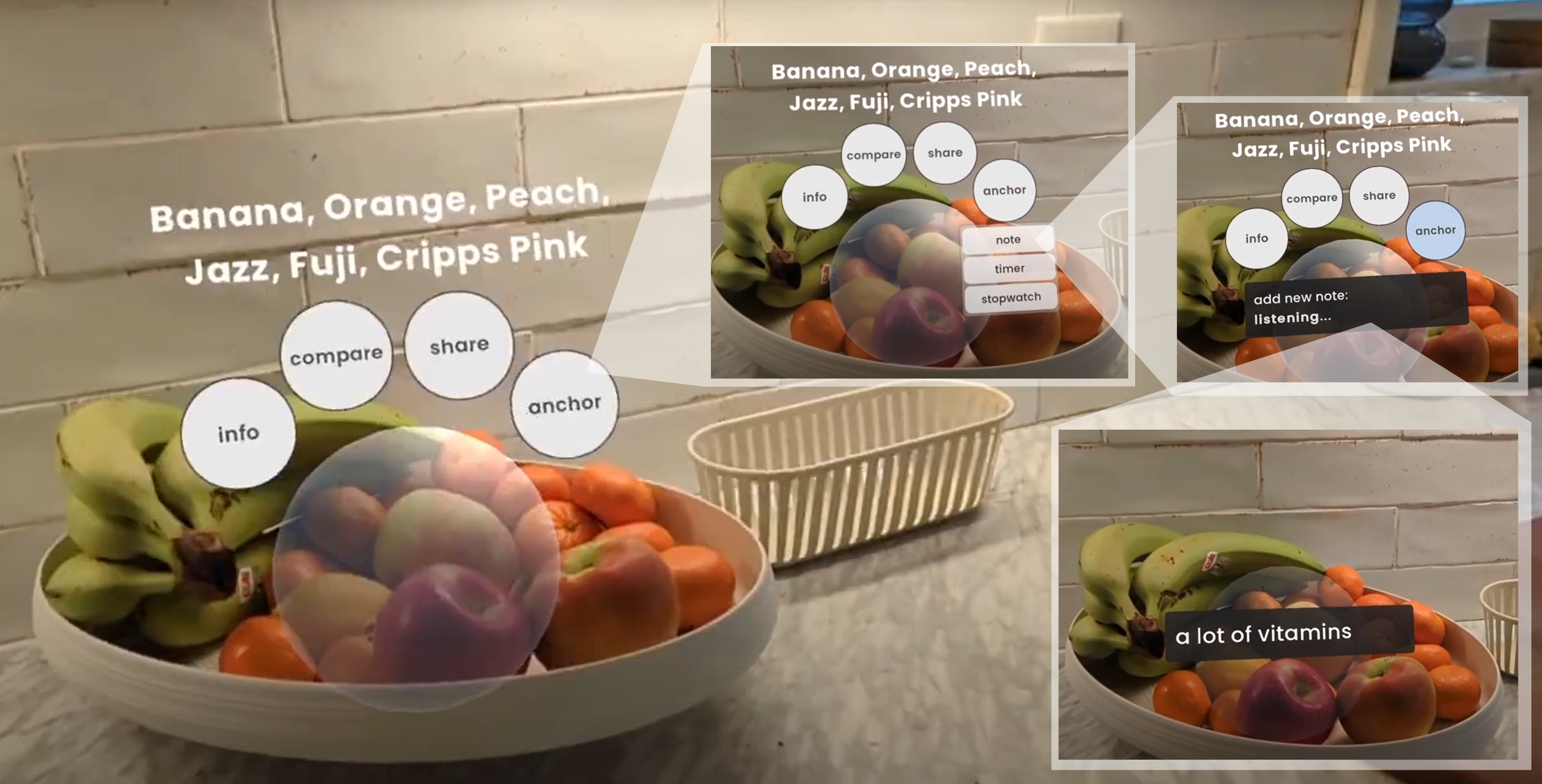}
  \caption{Example of anchoring contextual educational notes over real-world objects using \systemName.}
  \Description{The user anchors a spatial note on the fruits: "a lot of vitamins".}
  \label{fig:learning}
\end{figure}

\systemName~ can offer immersive or interactive learning experiences by augmenting physical objects with contextual educational content. By pointing their device at an object, users may access relevant explanations or demonstrations that enhance their understanding of the subject matter.
For instance, as illustrated in Figure~\ref{fig:learning}, \systemName~ could facilitate learning healthy eating habits for children. A child can point their device at a fruit bowl and instantly see information about the different fruits, such as their names, nutritional values, and the vitamins they contain.
\new{We also envision users leaving spatial notes on objects, and these could have changeable different visibility options. For instance, a user might leave a personal note about something they found out which they would like to attach as a reminder. They might further set the visibility to a specific group, e.g., their family members or coworkers, so that others can see and be aware of the new information.}

%This interactive learning experience can help children develop an understanding of the importance of a balanced diet and the benefits of various fruits.

% \subsection{Continued Contextual Interactions}

% The life of an intelligent tomato

\subsection{IoT Connectivity}

\systemName's  XR interface is complemented by a MLLM backend, unlike traditional IoT control interfaces that often limit interactions with devices to discrete apps. % or simplistic voice commands.  %, %facilitating a seamless blend of the physical and digital worlds.
Through its object tracking, \systemName~ could enable users to interact with their IoT devices in a spatial context to allow for real-time visual feedback and control within their immediate environment. 

\begin{figure}[b]
  \centering
  \includegraphics[width=1\linewidth]{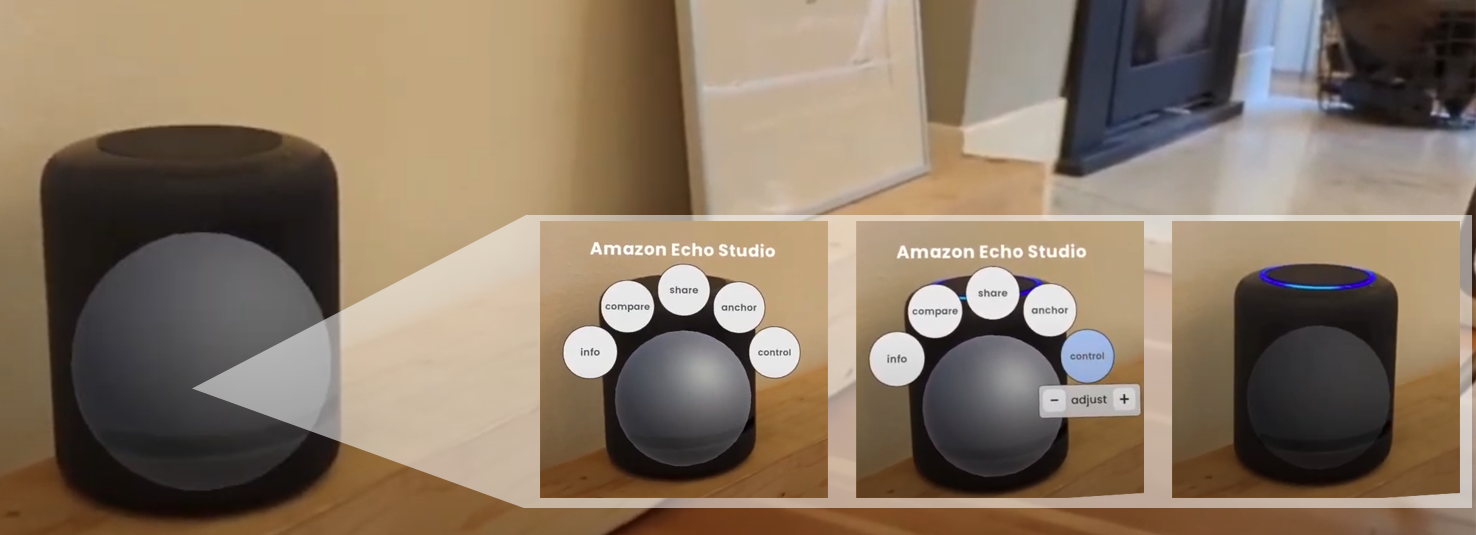}
  \caption{\systemName~ detects devices, such as speakers, and provides custom UI elements for IoT control.}
  \Description{The user controls the speaker using the AR interface.}
  \label{fig:iot}
\end{figure}

For example, in Figure~\ref{fig:iot}, the user controls their smart speaker using \systemName. They adjust the volume of the speaker using the touch UI or  language commands utilizing the MLLM backend. Such connectivity scenarios expand beyond speakers and could also be enabled on thermostats, smart lights, and other edge IoT devices.

\section{Discussion}
\label{discussion}

Our daily environment is ubiquitously augmented with various forms of annotations, from product packaging and price tags to traffic signs and personali post-it reminders. This ubiquitous augmentation, a rudimentary form of AR, is facilitated by %the advent of 
language, writing, and scalable printing technologies. It serves as a means to asynchronously convey contextual information, albeit in a static and limited manner, requiring manual interpretation and application by the user. Although machine-readable markers like barcodes, QR codes, and NFC tags have simplified certain interactions, they fall short in offering dynamic, object-relevant actionable insights.

\paragraph{Augmenting Objects with Intelligence}  
Advancements in computer vision and LLMs now enable devices to not only recognize generic object categories but also distinguish between individual instances based on their spatial context. This unlocks the potential for personalized, instance-specific interactions to transform objects into intelligent entities with their own "memories" of past interactions.  Accordingly, AOI has the potential to transition from the concept of smart tools to a reality where intelligence is an inherent characteristic of every object.
% , seamlessly integrating informative and actionable insights into our natural interaction flow with our surroundings.

\paragraph{Context-Aware Interactions}
\new{
While the list of actions we currently provide in our system is not exhaustive, we view it as a foundation that can be expanded by the community.
Looking ahead, we anticipate contributions from researchers and practitioners, e.g., integrating methods such as \textit{OmniActions}~\cite{li_omniactions_2024} for custom action suggestions, \textit{WorldGaze}~\cite{mayer_enhancing_2020} for gaze input, and \textit{GazePointAR}~\cite{lee_gazepointar_2024} for pronoun disambiguation in speech interactions.
}
%, to further enrich the system.
%Looking ahead, 
Using these extensions, \systemName~ could evolve to become even more attuned to the user's context (e.g. current location and activity, object relationships, persistence of object and data, see Figure~\ref{fig:applications}) to further customize actions and information based on historical interactions with objects in specific settings.
For instance, viewing a food item's packaging in a store might trigger suggestions for understanding its allergens and nutritional content, while the same item at home could offer cooking instructions and allow the user to directly set a digital cooking timer to the appropriate duration.
On certain smartphones, for instance,  \textit{Siri Suggestions}\footnote{Siri Suggestions on iPhone: \url{https://support.apple.com/guide/iphone/about-siri-suggestions-iph6f94af287/ios}} already offer context-aware recommendations. The functionality of this sort of context-awareness could be extended using AR interfaces.

\paragraph{\systemName~ on AR Headsets and Glasses}  
\new{
Participants expressed enthusiasm for \systemName~ on wearable AR. P3 mentioned the current challenges of typing on headsets and the advantages of voice and selection-based interfaces such as \systemName. P2 and P8 emphasized the practicality of using features like spatial timers on glasses for more glanceable interactions. They noted \systemName~ enhance the ease of comparing and selecting objects. P8 suggested potential cross-device interactions, such as starting a task on glasses and continuing on a phone for detailed comparisons.
}

\new{
As noted in Section~\ref{implementation}, a possible implementation on today's headsets is currently challenging due to restricted access to camera streams (\textit{Quest 3}) or limited FOV (\textit{HoloLens}).
As an alternative, future projects might explore attaching a webcam onto a \textit{Quest 3} and manually calibrating the camera for \systemName~ detection and interactions. Our open-source repository~\cite{google_xr-objects_2024} contains a \textit{Unity} guide for processing a webcam feed on headsets to implement this. 
We envision that on headsets, developers could use mid-air gestures, rays, or other hand-based interaction methods~\cite{bawa_building_2022}.
}

% Regarding always-on bubbles, P1 suggests “timers/notes [can be] hidden once attached, or showing only a small indicating icon – only becomes visible when time is about to be up / when I intend, e.g. when dwelling on the object.” We will add these to Discussion (§6).

\paragraph{LLM: Hallucinations and AI Improvements}
\new{
We recognize LLM as an emerging technology. In this work, our primary focus is enhancing spatial interaction, not solely LLM accuracy. As AI evolves, so will our system, incorporating user feedback to refine LLM interactions based on object-centric data.
}

\new{
In our study, participants appreciated \systemName ' reduced risk of hallucination due to its object-instance-based spatialization. P4 noted: “\textit{The control and feedback for intermediate steps – which object was recognized, what the model thinks it is – afforded by \systemName~ provides more confidence that the model isn't hallucinating and makes it easier to spot when it is.}”
}

\paragraph{Leveraging Emerging Artificial General Intelligence (AGI)} 
The integration of emerging AGIs~\cite{morris_levels_2024}, foreshadowed by models like \textit{Gemini} or \textit{GPT-4}, opens up new opportunities for autonomous problem-solving within XR environments. AGI's potential to dynamically generate user interfaces in response to user queries could transform the way we interact with our physical world.
For instance, the user could ask the system to visualize the nutritional values of a product in a pie chart, i.e., by generating the code to create the user interface and graphs on-the-fly without being pre-programmed to create the  chart. Going a step further, one may imagine AGI-driven systems that not only respond to user prompts but also proactively offer assistance surfaced through a new action in the context menu, such as assembling a set of \textit{Lego} blocks into a desired structure through real-time, augmented instructions.
%This scenario shows the transformative capability of \systemName~ to facilitate on-the-fly solutions to physical tasks within an augmented space.

\paragraph{Object Detection and Selection}
\new{
Our real-time object detection operates at 31 fps on a \textit{Galaxy S21} smartphone without needing pre-processing. MLLM queries respond in approximately three seconds.
Currently, the object selection is bound to the \textit{MediaPipe} classifier's output. We plan to explore enhancements like subregion selection, as P1 suggested: "\textit{would be nice [to] manipulate the scene with touch gestures [to] lead the AI to detect a thin object.}"
}

\paragraph{Linking Realities}  
As we adopt these new innovations in AI and XR, we are likely to see significant changes in how we interact with the physical objects around us~\cite{hirzle_when_2023, nebeling_designing_2024}. 
We envision a future where physical items no longer need conventional labels or tags, relying instead on AOI for context and interaction. The merging of digital and physical worlds might bring about new connections, like \textit{direct links between digital files and physical items}. This could start a new paradigm where the digital and physical worlds blend together smoothly, without clear boundaries.

\section{Conclusion}
In this paper, we introduced \paradigmName~ (AOI), a novel paradigm that seamlessly integrates digital capabilities into physical objects through the use of \systemName. Our prototype system demonstrates the potential of AOI to transform how users interact with their surroundings by leveraging advancements in computer vision, spatial understanding, and Multimodal -LLM. The results of our user study show that \systemName~ significantly outperforms traditional multimodal AI interfaces, with participants completing tasks an average of 24\% shorter time and reporting higher levels of satisfaction, ease of use, and perceived responsiveness. By enabling familiar, interactions with everyday objects through anchored AR content and natural language processing, \systemName~ paves the way for a future where the boundaries between the physical and digital worlds become increasingly blurred. As we continue to expand the capabilities of \systemName, we envision a wide range of applications spanning domains such as cooking, productivity, and connectivity, ultimately leading to a more engaging, efficient, and immersive way of interacting with our surroundings.

\bibliographystyle{ACM-Reference-Format}
\bibliography{XR-Objects}


\begin{thebibliography}{87}

%%% the Web address.

\ifx \showCODEN    \undefined \def \showCODEN     #1{\unskip}     \fi
\ifx \showDOI      \undefined \def \showDOI       #1{#1}\fi
\ifx \showISBNx    \undefined \def \showISBNx     #1{\unskip}     \fi
\ifx \showISBNxiii \undefined \def \showISBNxiii  #1{\unskip}     \fi
\ifx \showISSN     \undefined \def \showISSN      #1{\unskip}     \fi
\ifx \showLCCN     \undefined \def \showLCCN      #1{\unskip}     \fi
\ifx \shownote     \undefined \def \shownote      #1{#1}          \fi
\ifx \showarticletitle \undefined \def \showarticletitle #1{#1}   \fi
\ifx \showURL      \undefined \def \showURL       {\relax}        \fi
% invisible to TeX
\providecommand\bibfield[2]{#2}
\providecommand\bibinfo[2]{#2}
\providecommand\natexlab[1]{#1}
\providecommand\showeprint[2][]{arXiv:#2}

\bibitem[Ahuja et~al\mbox{.}(2019)]%
        {ahuja_lightanchors_2019}
\bibfield{author}{\bibinfo{person}{Karan Ahuja}, \bibinfo{person}{Sujeath
  Pareddy}, \bibinfo{person}{Robert Xiao}, \bibinfo{person}{Mayank Goel}, {and}
  \bibinfo{person}{Chris Harrison}.} \bibinfo{year}{2019}\natexlab{}.
\newblock \showarticletitle{{LightAnchors}: {Appropriating} {Point} {Lights}
  for {Spatially}-{Anchored} {Augmented} {Reality} {Interfaces}}. In
  \bibinfo{booktitle}{\emph{Proceedings of the 32nd {Annual} {ACM} {Symposium}
  on {User} {Interface} {Software} and {Technology}}}
  \emph{(\bibinfo{series}{{UIST} '19})}. \bibinfo{publisher}{Association for
  Computing Machinery}, \bibinfo{address}{New York, NY, USA},
  \bibinfo{pages}{189--196}.
\newblock
\showISBNx{978-1-4503-6816-2}
\urldef\tempurl%
\url{https://doi.org/10.1145/3332165.3347884}
\showDOI{\tempurl}


\bibitem[Airth(1993)]%
        {airth_navigation_1993}
\bibfield{author}{\bibinfo{person}{David~R Airth}.}
  \bibinfo{year}{1993}\natexlab{}.
\newblock \showarticletitle{Navigation in pop-up menus}. In
  \bibinfo{booktitle}{\emph{{INTERACT} '93 and {CHI} '93 {Conference}
  {Companion} on {Human} {Factors} in {Computing} {Systems}}}
  \emph{(\bibinfo{series}{{CHI} '93})}. \bibinfo{publisher}{Association for
  Computing Machinery}, \bibinfo{address}{New York, NY, USA},
  \bibinfo{pages}{115--116}.
\newblock
\showISBNx{978-0-89791-574-8}
\urldef\tempurl%
\url{https://doi.org/10.1145/259964.260139}
\showDOI{\tempurl}
\newblock
\shownote{event-place: Amsterdam, The Netherlands}.


\bibitem[Auda et~al\mbox{.}(2023)]%
        {auda_scoping_2023}
\bibfield{author}{\bibinfo{person}{Jonas Auda}, \bibinfo{person}{Uwe
  Gruenefeld}, \bibinfo{person}{Sarah Faltaous}, \bibinfo{person}{Sven Mayer},
  {and} \bibinfo{person}{Stefan Schneegass}.} \bibinfo{year}{2023}\natexlab{}.
\newblock \showarticletitle{A scoping survey on cross-reality systems}.
\newblock \bibinfo{journal}{\emph{Comput. Surveys}} \bibinfo{volume}{56},
  \bibinfo{number}{4} (\bibinfo{year}{2023}), \bibinfo{pages}{1--38}.
\newblock
\newblock
\shownote{Publisher: ACM New York, NY, USA}.


\bibitem[Banovic et~al\mbox{.}(2011)]%
        {banovic_design_2011}
\bibfield{author}{\bibinfo{person}{Nikola Banovic}, \bibinfo{person}{Frank
  Chun~Yat Li}, \bibinfo{person}{David Dearman}, \bibinfo{person}{Koji Yatani},
  {and} \bibinfo{person}{Khai~N Truong}.} \bibinfo{year}{2011}\natexlab{}.
\newblock \showarticletitle{Design of unimanual multi-finger pie menu
  interaction}. In \bibinfo{booktitle}{\emph{Proceedings of the {ACM}
  {International} {Conference} on {Interactive} {Tabletops} and {Surfaces}}}
  \emph{(\bibinfo{series}{{ITS} '11})}. \bibinfo{publisher}{Association for
  Computing Machinery}, \bibinfo{address}{New York, NY, USA},
  \bibinfo{pages}{120--129}.
\newblock
\showISBNx{978-1-4503-0871-7}
\urldef\tempurl%
\url{https://doi.org/10.1145/2076354.2076378}
\showDOI{\tempurl}
\newblock
\shownote{event-place: Kobe, Japan}.


\bibitem[Bawa(2022)]%
        {bawa_building_2022}
\bibfield{author}{\bibinfo{person}{Navyata Bawa}.}
  \bibinfo{year}{2022}\natexlab{}.
\newblock \bibinfo{title}{Building {Intuitive} {Interactions} in {VR}:
  {Interaction} {SDK}, {First} {Hand} {Showcase} and {Other} {Resources}}.
\newblock
\newblock
\urldef\tempurl%
\url{https://developers.facebook.com/blog/post/2022/11/22/building-intuitive-interactions-vr/}
\showURL{%
\tempurl}


\bibitem[Blackwell(2006)]%
        {blackwell_reification_2006}
\bibfield{author}{\bibinfo{person}{Alan~F Blackwell}.}
  \bibinfo{year}{2006}\natexlab{}.
\newblock \showarticletitle{The reification of metaphor as a design tool}.
\newblock \bibinfo{journal}{\emph{ACM Trans. Comput.-Hum. Interact.}}
  \bibinfo{volume}{13}, \bibinfo{number}{4} (\bibinfo{date}{Dec.}
  \bibinfo{year}{2006}), \bibinfo{pages}{490--530}.
\newblock
\showISSN{1073-0516}
\urldef\tempurl%
\url{https://doi.org/10.1145/1188816.1188820}
\showDOI{\tempurl}
\newblock
\shownote{Place: New York, NY, USA Publisher: Association for Computing
  Machinery}.


\bibitem[Bonnail et~al\mbox{.}(2023)]%
        {bonnail_memory_2023}
\bibfield{author}{\bibinfo{person}{Elise Bonnail}, \bibinfo{person}{Wen-Jie
  Tseng}, \bibinfo{person}{Mark Mcgill}, \bibinfo{person}{Eric Lecolinet},
  \bibinfo{person}{Samuel Huron}, {and} \bibinfo{person}{Jan Gugenheimer}.}
  \bibinfo{year}{2023}\natexlab{}.
\newblock \showarticletitle{Memory {Manipulations} in {Extended} {Reality}}. In
  \bibinfo{booktitle}{\emph{Proceedings of the 2023 {CHI} {Conference} on
  {Human} {Factors} in {Computing} {Systems}}} \emph{(\bibinfo{series}{{CHI}
  '23})}. \bibinfo{publisher}{Association for Computing Machinery},
  \bibinfo{address}{New York, NY, USA}.
\newblock
\showISBNx{978-1-4503-9421-5}
\newblock
\shownote{event-place: Hamburg,Germany}.


\bibitem[Caetano and Sra(2022)]%
        {caetano_arfy_2022}
\bibfield{author}{\bibinfo{person}{Arthur Caetano} {and} \bibinfo{person}{Misha
  Sra}.} \bibinfo{year}{2022}\natexlab{}.
\newblock \showarticletitle{{ARfy}: {A} {Pipeline} for {Adapting} {3D} {Scenes}
  to {Augmented} {Reality}}. In \bibinfo{booktitle}{\emph{Adjunct {Proceedings}
  of the 35th {Annual} {ACM} {Symposium} on {User} {Interface} {Software} and
  {Technology}}} \emph{(\bibinfo{series}{{UIST} '22 {Adjunct}})}.
  \bibinfo{publisher}{Association for Computing Machinery},
  \bibinfo{address}{New York, NY, USA}, \bibinfo{pages}{1--3}.
\newblock
\showISBNx{978-1-4503-9321-8}
\urldef\tempurl%
\url{https://doi.org/10.1145/3526114.3558697}
\showDOI{\tempurl}


\bibitem[Campos~Zamora et~al\mbox{.}(2024)]%
        {campos_zamora_moirewidgets_2024}
\bibfield{author}{\bibinfo{person}{Daniel Campos~Zamora},
  \bibinfo{person}{Mustafa~Doga Dogan}, \bibinfo{person}{Alexa~F Siu},
  \bibinfo{person}{Eunyee Koh}, {and} \bibinfo{person}{Chang Xiao}.}
  \bibinfo{year}{2024}\natexlab{}.
\newblock \showarticletitle{{MoiréWidgets}: {High}-{Precision}, {Passive}
  {Tangible} {Interfaces} via {Moiré} {Effect}}. In
  \bibinfo{booktitle}{\emph{Proceedings of the {CHI} {Conference} on {Human}
  {Factors} in {Computing} {Systems}}}. \bibinfo{publisher}{ACM},
  \bibinfo{address}{Honolulu HI USA}, \bibinfo{pages}{1--10}.
\newblock
\showISBNx{9798400703300}
\urldef\tempurl%
\url{https://doi.org/10.1145/3613904.3642734}
\showDOI{\tempurl}


\bibitem[Chen et~al\mbox{.}(2023)]%
        {chen_pali_2023}
\bibfield{author}{\bibinfo{person}{Xi Chen}, \bibinfo{person}{Xiao Wang},
  \bibinfo{person}{Soravit Changpinyo}, \bibinfo{person}{A.~J. Piergiovanni},
  \bibinfo{person}{Piotr Padlewski}, \bibinfo{person}{Daniel Salz},
  \bibinfo{person}{Sebastian Goodman}, \bibinfo{person}{Adam Grycner},
  \bibinfo{person}{Basil Mustafa}, \bibinfo{person}{Lucas Beyer},
  \bibinfo{person}{Alexander Kolesnikov}, \bibinfo{person}{Joan Puigcerver},
  \bibinfo{person}{Nan Ding}, \bibinfo{person}{Keran Rong},
  \bibinfo{person}{Hassan Akbari}, \bibinfo{person}{Gaurav Mishra},
  \bibinfo{person}{Linting Xue}, \bibinfo{person}{Ashish Thapliyal},
  \bibinfo{person}{James Bradbury}, \bibinfo{person}{Weicheng Kuo},
  \bibinfo{person}{Mojtaba Seyedhosseini}, \bibinfo{person}{Chao Jia},
  \bibinfo{person}{Burcu~Karagol Ayan}, \bibinfo{person}{Carlos Riquelme},
  \bibinfo{person}{Andreas Steiner}, \bibinfo{person}{Anelia Angelova},
  \bibinfo{person}{Xiaohua Zhai}, \bibinfo{person}{Neil Houlsby}, {and}
  \bibinfo{person}{Radu Soricut}.} \bibinfo{year}{2023}\natexlab{}.
\newblock \bibinfo{title}{{PaLI}: {A} {Jointly}-{Scaled} {Multilingual}
  {Language}-{Image} {Model}}.
\newblock
\newblock
\newblock
\shownote{\_eprint: 2209.06794}.


\bibitem[Chen et~al\mbox{.}(2020)]%
        {chen_augmenting_2020}
\bibfield{author}{\bibinfo{person}{Zhutian Chen}, \bibinfo{person}{Wai Tong},
  \bibinfo{person}{Qianwen Wang}, \bibinfo{person}{Benjamin Bach}, {and}
  \bibinfo{person}{Huamin Qu}.} \bibinfo{year}{2020}\natexlab{}.
\newblock \showarticletitle{Augmenting {Static} {Visualizations} with
  {PapARVis} {Designer}}. In \bibinfo{booktitle}{\emph{Proceedings of the 2020
  {CHI} {Conference} on {Human} {Factors} in {Computing} {Systems}}}
  \emph{(\bibinfo{series}{{CHI} '20})}. \bibinfo{publisher}{Association for
  Computing Machinery}, \bibinfo{address}{New York, NY, USA},
  \bibinfo{pages}{1--12}.
\newblock
\showISBNx{978-1-4503-6708-0}
\urldef\tempurl%
\url{https://doi.org/10.1145/3313831.3376436}
\showDOI{\tempurl}


\bibitem[Cheng et~al\mbox{.}(2021)]%
        {cheng_semanticadapt_2021}
\bibfield{author}{\bibinfo{person}{Yifei Cheng}, \bibinfo{person}{Yukang Yan},
  \bibinfo{person}{Xin Yi}, \bibinfo{person}{Yuanchun Shi}, {and}
  \bibinfo{person}{David Lindlbauer}.} \bibinfo{year}{2021}\natexlab{}.
\newblock \showarticletitle{{SemanticAdapt}: {Optimization}-based {Adaptation}
  of {Mixed} {Reality} {Layouts} {Leveraging} {Virtual}-{Physical} {Semantic}
  {Connections}}. In \bibinfo{booktitle}{\emph{The 34th {Annual} {ACM}
  {Symposium} on {User} {Interface} {Software} and {Technology}}}
  \emph{(\bibinfo{series}{{UIST} '21})}. \bibinfo{publisher}{Association for
  Computing Machinery}, \bibinfo{address}{New York, NY, USA},
  \bibinfo{pages}{282--297}.
\newblock
\showISBNx{978-1-4503-8635-7}
\urldef\tempurl%
\url{https://doi.org/10.1145/3472749.3474750}
\showDOI{\tempurl}
\newblock
\shownote{event-place: Virtual Event, USA}.


\bibitem[Cheng et~al\mbox{.}(2023)]%
        {cheng_interactionadapt_2023}
\bibfield{author}{\bibinfo{person}{Yi~Fei Cheng}, \bibinfo{person}{Christoph
  Gebhardt}, {and} \bibinfo{person}{Christian Holz}.}
  \bibinfo{year}{2023}\natexlab{}.
\newblock \showarticletitle{{InteractionAdapt}: {Interaction}-driven
  {Workspace} {Adaptation} for {Situated} {Virtual} {Reality} {Environments}}.
  In \bibinfo{booktitle}{\emph{Proceedings of the 36th {Annual} {ACM}
  {Symposium} on {User} {Interface} {Software} and {Technology}}}
  \emph{(\bibinfo{series}{{UIST} '23})}. \bibinfo{publisher}{Association for
  Computing Machinery}, \bibinfo{address}{New York, NY, USA}.
\newblock
\showISBNx{9798400701320}
\newblock
\shownote{event-place: San Francisco, CA,USA}.


\bibitem[Chidambaram et~al\mbox{.}(2021)]%
        {chidambaram_processar_2021}
\bibfield{author}{\bibinfo{person}{Subramanian Chidambaram},
  \bibinfo{person}{Hank Huang}, \bibinfo{person}{Fengming He},
  \bibinfo{person}{Xun Qian}, \bibinfo{person}{Ana~M Villanueva},
  \bibinfo{person}{Thomas~S Redick}, \bibinfo{person}{Wolfgang Stuerzlinger},
  {and} \bibinfo{person}{Karthik Ramani}.} \bibinfo{year}{2021}\natexlab{}.
\newblock \showarticletitle{{ProcessAR}: {An} augmented reality-based tool to
  create in-situ procedural {2D}/{3D} {AR} {Instructions}}. In
  \bibinfo{booktitle}{\emph{Proceedings of the 2021 {ACM} {Designing}
  {Interactive} {Systems} {Conference}}} \emph{(\bibinfo{series}{{DIS} '21})}.
  \bibinfo{publisher}{Association for Computing Machinery},
  \bibinfo{address}{New York, NY, USA}, \bibinfo{pages}{234--249}.
\newblock
\showISBNx{978-1-4503-8476-6}
\urldef\tempurl%
\url{https://doi.org/10.1145/3461778.3462126}
\showDOI{\tempurl}


\bibitem[Chidambaram et~al\mbox{.}(2022)]%
        {chidambaram_editar_2022}
\bibfield{author}{\bibinfo{person}{Subramanian Chidambaram},
  \bibinfo{person}{Sai~Swarup Reddy}, \bibinfo{person}{Matthew Rumple},
  \bibinfo{person}{Ananya Ipsita}, \bibinfo{person}{Ana Villanueva},
  \bibinfo{person}{Thomas Redick}, \bibinfo{person}{Wolfgang Stuerzlinger},
  {and} \bibinfo{person}{Karthik Ramani}.} \bibinfo{year}{2022}\natexlab{}.
\newblock \showarticletitle{{EditAR}: {A} {Digital} {Twin} {Authoring}
  {Environment} for {Creation} of {AR}/{VR} and {Video} {Instructions} from a
  {Single} {Demonstration}}. In \bibinfo{booktitle}{\emph{2022 {IEEE}
  {International} {Symposium} on {Mixed} and {Augmented} {Reality} ({ISMAR})}}.
  \bibinfo{publisher}{IEEE}, \bibinfo{address}{Singapore, Singapore},
  \bibinfo{pages}{326--335}.
\newblock
\showISBNx{978-1-66545-325-7}
\urldef\tempurl%
\url{https://doi.org/10.1109/ISMAR55827.2022.00048}
\showDOI{\tempurl}


\bibitem[Chulpongsatorn et~al\mbox{.}(2023a)]%
        {chulpongsatorn_augmented_2023}
\bibfield{author}{\bibinfo{person}{Neil Chulpongsatorn},
  \bibinfo{person}{Mille~Skovhus Lunding}, \bibinfo{person}{Nishan Soni}, {and}
  \bibinfo{person}{Ryo Suzuki}.} \bibinfo{year}{2023}\natexlab{a}.
\newblock \showarticletitle{Augmented {Math}: {Authoring} {AR}-{Based}
  {Explorable} {Explanations} by {Augmenting} {Static} {Math} {Textbooks}}.
\newblock  (\bibinfo{date}{July} \bibinfo{year}{2023}).
\newblock
\urldef\tempurl%
\url{https://doi.org/10.1145/3586183.3606827}
\showDOI{\tempurl}
\newblock
\shownote{\_eprint: 2307.16112}.


\bibitem[Chulpongsatorn et~al\mbox{.}(2023b)]%
        {chulpongsatorn_holotouch_2023}
\bibfield{author}{\bibinfo{person}{Neil Chulpongsatorn},
  \bibinfo{person}{Wesley Willett}, {and} \bibinfo{person}{Ryo Suzuki}.}
  \bibinfo{year}{2023}\natexlab{b}.
\newblock \showarticletitle{{HoloTouch}: {Interacting} with {Mixed} {Reality}
  {Visualizations} {Through} {Smartphone} {Proxies}}.
\newblock  (\bibinfo{date}{March} \bibinfo{year}{2023}).
\newblock
\urldef\tempurl%
\url{https://doi.org/10.1145/3544549.3585738}
\showDOI{\tempurl}
\newblock
\shownote{\_eprint: 2303.08916}.


\bibitem[{Davison}(2003)]%
        {davison_real-time_2003}
\bibfield{author}{\bibinfo{person}{{Davison}}.}
  \bibinfo{year}{2003}\natexlab{}.
\newblock \showarticletitle{Real-time simultaneous localisation and mapping
  with a single camera}. In \bibinfo{booktitle}{\emph{Proceedings {Ninth}
  {IEEE} {International} {Conference} on {Computer} {Vision}}}.
  \bibinfo{publisher}{IEEE}, \bibinfo{pages}{1403--1410}.
\newblock


\bibitem[Dogan(2024)]%
        {dogan_ubiquitous_2024}
\bibfield{author}{\bibinfo{person}{Mustafa~Doga Dogan}.}
  \bibinfo{year}{2024}\natexlab{}.
\newblock \bibinfo{title}{Ubiquitous {Metadata}: {Design} and {Fabrication} of
  {Embedded} {Markers} for {Real}-{World} {Object} {Identification} and
  {Interaction}}.
\newblock
\newblock
\urldef\tempurl%
\url{https://doi.org/10.48550/arXiv.2407.11748}
\showDOI{\tempurl}
\newblock
\shownote{arXiv:2407.11748 [cs]}.


\bibitem[Dogan et~al\mbox{.}(2022a)]%
        {dogan_fabricate_2022}
\bibfield{author}{\bibinfo{person}{Mustafa~Doga Dogan},
  \bibinfo{person}{Patrick Baudisch}, \bibinfo{person}{Hrvoje Benko},
  \bibinfo{person}{Michael Nebeling}, \bibinfo{person}{Huaishu Peng},
  \bibinfo{person}{Valkyrie Savage}, {and} \bibinfo{person}{Stefanie Mueller}.}
  \bibinfo{year}{2022}\natexlab{a}.
\newblock \showarticletitle{Fabricate {It} or {Render} {It}? {Digital}
  {Fabrication} vs. {Virtual} {Reality} for {Creating} {Objects} {Instantly}}.
  In \bibinfo{booktitle}{\emph{{CHI} {Conference} on {Human} {Factors} in
  {Computing} {Systems} {Extended} {Abstracts}}}. \bibinfo{publisher}{ACM},
  \bibinfo{address}{New Orleans LA USA}, \bibinfo{pages}{1--4}.
\newblock
\showISBNx{978-1-4503-9156-6}
\urldef\tempurl%
\url{https://doi.org/10.1145/3491101.3516510}
\showDOI{\tempurl}


\bibitem[Dogan et~al\mbox{.}(2021)]%
        {dogan_sensicut_2021}
\bibfield{author}{\bibinfo{person}{Mustafa~Doga Dogan}, \bibinfo{person}{Steven
  Vidal~Acevedo Colon}, \bibinfo{person}{Varnika Sinha}, \bibinfo{person}{Kaan
  Akşit}, {and} \bibinfo{person}{Stefanie Mueller}.}
  \bibinfo{year}{2021}\natexlab{}.
\newblock \showarticletitle{{SensiCut}: {Material}-{Aware} {Laser} {Cutting}
  {Using} {Speckle} {Sensing} and {Deep} {Learning}}. In
  \bibinfo{booktitle}{\emph{Proceedings of the 34th {Annual} {ACM} {Symposium}
  on {User} {Interface} {Software} and {Technology}}}.
  \bibinfo{publisher}{ACM}, \bibinfo{address}{Virtual Event USA},
  \bibinfo{pages}{15}.
\newblock
\showISBNx{978-1-4503-8635-7}
\urldef\tempurl%
\url{https://doi.org/10.1145/3472749.3474733}
\showDOI{\tempurl}


\bibitem[Dogan et~al\mbox{.}(2020)]%
        {dogan_g-id_2020}
\bibfield{author}{\bibinfo{person}{Mustafa~Doga Dogan}, \bibinfo{person}{Faraz
  Faruqi}, \bibinfo{person}{Andrew~Day Churchill}, \bibinfo{person}{Kenneth
  Friedman}, \bibinfo{person}{Leon Cheng}, \bibinfo{person}{Sriram
  Subramanian}, {and} \bibinfo{person}{Stefanie Mueller}.}
  \bibinfo{year}{2020}\natexlab{}.
\newblock \showarticletitle{G-{ID}: {Identifying} {3D} {Prints} {Using}
  {Slicing} {Parameters}}. In \bibinfo{booktitle}{\emph{Proceedings of the 2020
  {CHI} {Conference} on {Human} {Factors} in {Computing} {Systems}}}
  \emph{(\bibinfo{series}{{CHI} '20})}. \bibinfo{publisher}{Association for
  Computing Machinery}, \bibinfo{address}{New York, NY, USA},
  \bibinfo{pages}{1--13}.
\newblock
\showISBNx{978-1-4503-6708-0}
\urldef\tempurl%
\url{https://doi.org/10.1145/3313831.3376202}
\showDOI{\tempurl}


\bibitem[Dogan et~al\mbox{.}(2023a)]%
        {dogan_brightmarker_2023}
\bibfield{author}{\bibinfo{person}{Mustafa~Doga Dogan}, \bibinfo{person}{Raul
  Garcia-Martin}, \bibinfo{person}{Patrick~William Haertel},
  \bibinfo{person}{Jamison~John O'Keefe}, \bibinfo{person}{Ahmad Taka},
  \bibinfo{person}{Akarsh Aurora}, \bibinfo{person}{Raul Sanchez-Reillo}, {and}
  \bibinfo{person}{Stefanie Mueller}.} \bibinfo{year}{2023}\natexlab{a}.
\newblock \showarticletitle{{BrightMarker}: {3D} {Printed} {Fluorescent}
  {Markers} for {Object} {Tracking}}. In \bibinfo{booktitle}{\emph{Proceedings
  of the 36th {Annual} {ACM} {Symposium} on {User} {Interface} {Software} and
  {Technology}}} \emph{(\bibinfo{series}{{UIST} '23})}.
  \bibinfo{publisher}{Association for Computing Machinery},
  \bibinfo{address}{New York, NY, USA}, \bibinfo{pages}{1--13}.
\newblock
\showISBNx{9798400701320}
\urldef\tempurl%
\url{https://doi.org/10.1145/3586183.3606758}
\showDOI{\tempurl}


\bibitem[Dogan et~al\mbox{.}(2023b)]%
        {dogan_standarone_2023}
\bibfield{author}{\bibinfo{person}{Mustafa~Doga Dogan},
  \bibinfo{person}{Alexa~F. Siu}, \bibinfo{person}{Jennifer Healey},
  \bibinfo{person}{Curtis Wigington}, \bibinfo{person}{Chang Xiao}, {and}
  \bibinfo{person}{Tong Sun}.} \bibinfo{year}{2023}\natexlab{b}.
\newblock \showarticletitle{{StandARone}: {Infrared}-{Watermarked} {Documents}
  as {Portable} {Containers} of {AR} {Interaction} and {Personalization}}. In
  \bibinfo{booktitle}{\emph{Extended {Abstracts} of the 2023 {CHI} {Conference}
  on {Human} {Factors} in {Computing} {Systems}}} \emph{(\bibinfo{series}{{CHI}
  {EA} '23})}. \bibinfo{publisher}{Association for Computing Machinery},
  \bibinfo{address}{New York, NY, USA}, \bibinfo{pages}{1--7}.
\newblock
\showISBNx{978-1-4503-9422-2}
\urldef\tempurl%
\url{https://doi.org/10.1145/3544549.3585905}
\showDOI{\tempurl}


\bibitem[Dogan et~al\mbox{.}(2022b)]%
        {dogan_infraredtags_2022}
\bibfield{author}{\bibinfo{person}{Mustafa~Doga Dogan}, \bibinfo{person}{Ahmad
  Taka}, \bibinfo{person}{Michael Lu}, \bibinfo{person}{Yunyi Zhu},
  \bibinfo{person}{Akshat Kumar}, \bibinfo{person}{Aakar Gupta}, {and}
  \bibinfo{person}{Stefanie Mueller}.} \bibinfo{year}{2022}\natexlab{b}.
\newblock \showarticletitle{{InfraredTags}: {Embedding} {Invisible} {AR}
  {Markers} and {Barcodes} {Using} {Low}-{Cost}, {Infrared}-{Based} {3D}
  {Printing} and {Imaging} {Tools}}. In \bibinfo{booktitle}{\emph{Proceedings
  of the 2022 {CHI} {Conference} on {Human} {Factors} in {Computing}
  {Systems}}} \emph{(\bibinfo{series}{{CHI} '22})}.
  \bibinfo{publisher}{Association for Computing Machinery},
  \bibinfo{address}{New York, NY, USA}, \bibinfo{pages}{1--12}.
\newblock
\showISBNx{978-1-4503-9157-3}
\urldef\tempurl%
\url{https://doi.org/10.1145/3491102.3501951}
\showDOI{\tempurl}
\newblock
\shownote{Issue: Article 269 event-place: New Orleans, LA, USA}.


\bibitem[Dogan et~al\mbox{.}(2022c)]%
        {dogan_demonstrating_2022}
\bibfield{author}{\bibinfo{person}{Mustafa~Doga Dogan},
  \bibinfo{person}{Veerapatr Yotamornsunthorn}, \bibinfo{person}{Ahmad Taka},
  \bibinfo{person}{Yunyi Zhu}, \bibinfo{person}{Aakar Gupta}, {and}
  \bibinfo{person}{Stefanie Mueller}.} \bibinfo{year}{2022}\natexlab{c}.
\newblock \showarticletitle{Demonstrating {InfraredTags}: {Decoding}
  {Invisible} {3D} {Printed} {Tags} with {Convolutional} {Neural} {Networks}}.
  In \bibinfo{booktitle}{\emph{Extended {Abstracts} of the 2022 {CHI}
  {Conference} on {Human} {Factors} in {Computing} {Systems}}}
  \emph{(\bibinfo{series}{{CHI} {EA} '22})}. \bibinfo{publisher}{Association
  for Computing Machinery}, \bibinfo{address}{New York, NY, USA},
  \bibinfo{pages}{1--5}.
\newblock
\showISBNx{978-1-4503-9156-6}
\urldef\tempurl%
\url{https://doi.org/10.1145/3491101.3519905}
\showDOI{\tempurl}
\newblock
\shownote{Issue: Article 185 event-place: New Orleans, LA, USA}.


\bibitem[Du et~al\mbox{.}(2022)]%
        {du_opportunistic_2022}
\bibfield{author}{\bibinfo{person}{Ruofei Du}, \bibinfo{person}{Alex Olwal},
  \bibinfo{person}{Mathieu Le~Goc}, \bibinfo{person}{Shengzhi Wu},
  \bibinfo{person}{Danhang Tang}, \bibinfo{person}{Yinda Zhang},
  \bibinfo{person}{Jun Zhang}, \bibinfo{person}{David~Joseph Tan},
  \bibinfo{person}{Federico Tombari}, {and} \bibinfo{person}{David Kim}.}
  \bibinfo{year}{2022}\natexlab{}.
\newblock \showarticletitle{Opportunistic {Interfaces} for {Augmented}
  {Reality}: {Transforming} {Everyday} {Objects} into {Tangible} {6DoF}
  {Interfaces} {Using} {Ad} hoc {UI}}. In \bibinfo{booktitle}{\emph{Extended
  {Abstracts} of the 2022 {CHI} {Conference} on {Human} {Factors} in
  {Computing} {Systems}}} \emph{(\bibinfo{series}{{CHI} {EA} '22})}.
  \bibinfo{publisher}{Association for Computing Machinery},
  \bibinfo{address}{New York, NY, USA}, \bibinfo{pages}{1--4}.
\newblock
\showISBNx{978-1-4503-9156-6}
\urldef\tempurl%
\url{https://doi.org/10.1145/3491101.3519911}
\showDOI{\tempurl}


\bibitem[Du et~al\mbox{.}(2020)]%
        {du_depthlab_2020}
\bibfield{author}{\bibinfo{person}{Ruofei Du}, \bibinfo{person}{Eric Turner},
  \bibinfo{person}{Maksym Dzitsiuk}, \bibinfo{person}{Luca Prasso},
  \bibinfo{person}{Ivo Duarte}, \bibinfo{person}{Jason Dourgarian},
  \bibinfo{person}{Joao Afonso}, \bibinfo{person}{Jose Pascoal},
  \bibinfo{person}{Josh Gladstone}, \bibinfo{person}{Nuno Cruces}, {and}
  \bibinfo{person}{{others}}.} \bibinfo{year}{2020}\natexlab{}.
\newblock \showarticletitle{{DepthLab}: {Real}-time {3D} interaction with depth
  maps for mobile augmented reality}. In \bibinfo{booktitle}{\emph{Proceedings
  of the 33rd {Annual} {ACM} {Symposium} on {User} {Interface} {Software} and
  {Technology}}}. \bibinfo{pages}{829--843}.
\newblock


\bibitem[Fender and Holz(2022)]%
        {fender_causality-preserving_2022}
\bibfield{author}{\bibinfo{person}{Andreas~Rene Fender} {and}
  \bibinfo{person}{Christian Holz}.} \bibinfo{year}{2022}\natexlab{}.
\newblock \showarticletitle{Causality-preserving {Asynchronous} {Reality}}. In
  \bibinfo{booktitle}{\emph{Proceedings of the 2022 {CHI} {Conference} on
  {Human} {Factors} in {Computing} {Systems}}} \emph{(\bibinfo{series}{{CHI}
  '22})}. \bibinfo{publisher}{Association for Computing Machinery},
  \bibinfo{address}{New York, NY, USA}.
\newblock
\showISBNx{978-1-4503-9157-3}
\newblock
\shownote{event-place: New Orleans, LA,USA}.


\bibitem[Gonzalez et~al\mbox{.}(2024)]%
        {gonzalez_xdtk_2024}
\bibfield{author}{\bibinfo{person}{Eric~J. Gonzalez}, \bibinfo{person}{Khushman
  Patel}, \bibinfo{person}{Karan Ahuja}, {and} \bibinfo{person}{Mar
  Gonzalez-Franco}.} \bibinfo{year}{2024}\natexlab{}.
\newblock \showarticletitle{{XDTK}: {A} {Cross}-{Device} {Toolkit} for {Input}
  \& {Interaction} in {XR}}. In \bibinfo{booktitle}{\emph{2024 {IEEE}
  {Conference} on {Virtual} {Reality} and {3D} {User} {Interfaces} {Abstracts}
  and {Workshops} ({VRW})}}. \bibinfo{publisher}{IEEE}.
\newblock


\bibitem[Gonzalez-Franco and Colaco(2024)]%
        {gonzalez-franco_guidelines_2024}
\bibfield{author}{\bibinfo{person}{Mar Gonzalez-Franco} {and}
  \bibinfo{person}{Andrea Colaco}.} \bibinfo{year}{2024}\natexlab{}.
\newblock \showarticletitle{Guidelines for {Productivity} in {Virtual}
  {Reality}}.
\newblock \bibinfo{journal}{\emph{ACM Interactions Magazine}}
  (\bibinfo{year}{2024}).
\newblock
\newblock
\shownote{Publisher: ACM}.


\bibitem[Gonzalez-Franco and Lanier(2017)]%
        {gonzalez-franco_model_2017}
\bibfield{author}{\bibinfo{person}{Mar Gonzalez-Franco} {and}
  \bibinfo{person}{Jaron Lanier}.} \bibinfo{year}{2017}\natexlab{}.
\newblock \showarticletitle{Model of illusions and virtual reality}.
\newblock \bibinfo{journal}{\emph{Frontiers in psychology}}
  \bibinfo{volume}{8} (\bibinfo{year}{2017}), \bibinfo{pages}{273943}.
\newblock
\newblock
\shownote{Publisher: Frontiers}.


\bibitem[Google(2024a)]%
        {google_arcore_2024}
\bibfield{author}{\bibinfo{person}{Google}.} \bibinfo{year}{2024}\natexlab{a}.
\newblock \bibinfo{title}{{ARCore}}.
\newblock
\newblock
\urldef\tempurl%
\url{https://developers.google.com/ar}
\showURL{%
\tempurl}


\bibitem[Google(2024b)]%
        {google_xr-objects_2024}
\bibfield{author}{\bibinfo{person}{Google}.} \bibinfo{year}{2024}\natexlab{b}.
\newblock \bibinfo{title}{{XR}-{Objects} repository}.
\newblock
\newblock
\urldef\tempurl%
\url{https://github.com/google/xr-objects}
\showURL{%
\tempurl}
\newblock
\shownote{original-date: 2024-02-17T11:08:33Z}.


\bibitem[Han et~al\mbox{.}(2023)]%
        {han_blendmr_2023}
\bibfield{author}{\bibinfo{person}{Violet~Yinuo Han}, \bibinfo{person}{Hyunsung
  Cho}, \bibinfo{person}{Kiyosu Maeda}, \bibinfo{person}{Alexandra Ion}, {and}
  \bibinfo{person}{David Lindlbauer}.} \bibinfo{year}{2023}\natexlab{}.
\newblock \showarticletitle{{BlendMR}: {A} {Computational} {Method} to {Create}
  {Ambient} {Mixed} {Reality} {Interfaces}}.
\newblock \bibinfo{journal}{\emph{Proceedings of the ACM on Human-Computer
  Interaction}} \bibinfo{volume}{7}, \bibinfo{number}{ISS}
  (\bibinfo{date}{Nov.} \bibinfo{year}{2023}),
  \bibinfo{pages}{436:217--436:241}.
\newblock
\urldef\tempurl%
\url{https://doi.org/10.1145/3626472}
\showDOI{\tempurl}


\bibitem[Hartmann et~al\mbox{.}(2019)]%
        {hartmann_realitycheck_2019}
\bibfield{author}{\bibinfo{person}{Jeremy Hartmann}, \bibinfo{person}{Christian
  Holz}, \bibinfo{person}{Eyal Ofek}, {and} \bibinfo{person}{Andrew~D Wilson}.}
  \bibinfo{year}{2019}\natexlab{}.
\newblock \showarticletitle{Realitycheck: {Blending} virtual environments with
  situated physical reality}. In \bibinfo{booktitle}{\emph{Proceedings of the
  2019 {CHI} {Conference} on {Human} {Factors} in {Computing} {Systems}}}.
  \bibinfo{pages}{1--12}.
\newblock


\bibitem[He et~al\mbox{.}(2023)]%
        {he_ubi_2023}
\bibfield{author}{\bibinfo{person}{Fengming He}, \bibinfo{person}{Xiyun Hu},
  \bibinfo{person}{Jingyu Shi}, \bibinfo{person}{Xun Qian},
  \bibinfo{person}{Tianyi Wang}, {and} \bibinfo{person}{Karthik Ramani}.}
  \bibinfo{year}{2023}\natexlab{}.
\newblock \showarticletitle{Ubi {Edge}: {Authoring} {Edge}-{Based}
  {Opportunistic} {Tangible} {User} {Interfaces} in {Augmented} {Reality}}. In
  \bibinfo{booktitle}{\emph{Proceedings of the 2023 {CHI} {Conference} on
  {Human} {Factors} in {Computing} {Systems}}} \emph{(\bibinfo{series}{{CHI}
  '23})}. \bibinfo{publisher}{Association for Computing Machinery},
  \bibinfo{address}{New York, NY, USA}, \bibinfo{pages}{1--14}.
\newblock
\showISBNx{978-1-4503-9421-5}
\urldef\tempurl%
\url{https://doi.org/10.1145/3544548.3580704}
\showDOI{\tempurl}


\bibitem[Henderson and Feiner(2008)]%
        {henderson_opportunistic_2008}
\bibfield{author}{\bibinfo{person}{Steven~J. Henderson} {and}
  \bibinfo{person}{Steven Feiner}.} \bibinfo{year}{2008}\natexlab{}.
\newblock \showarticletitle{Opportunistic controls: leveraging natural
  affordances as tangible user interfaces for augmented reality}. In
  \bibinfo{booktitle}{\emph{Proceedings of the 2008 {ACM} symposium on
  {Virtual} reality software and technology}} \emph{(\bibinfo{series}{{VRST}
  '08})}. \bibinfo{publisher}{Association for Computing Machinery},
  \bibinfo{address}{New York, NY, USA}, \bibinfo{pages}{211--218}.
\newblock
\showISBNx{978-1-59593-951-7}
\urldef\tempurl%
\url{https://doi.org/10.1145/1450579.1450625}
\showDOI{\tempurl}


\bibitem[Hettiarachchi and Wigdor(2016)]%
        {hettiarachchi_annexing_2016}
\bibfield{author}{\bibinfo{person}{Anuruddha Hettiarachchi} {and}
  \bibinfo{person}{Daniel Wigdor}.} \bibinfo{year}{2016}\natexlab{}.
\newblock \showarticletitle{Annexing {Reality}: {Enabling} {Opportunistic}
  {Use} of {Everyday} {Objects} as {Tangible} {Proxies} in {Augmented}
  {Reality}}. In \bibinfo{booktitle}{\emph{Proceedings of the 2016 {CHI}
  {Conference} on {Human} {Factors} in {Computing} {Systems}}}
  \emph{(\bibinfo{series}{{CHI} '16})}. \bibinfo{publisher}{Association for
  Computing Machinery}, \bibinfo{address}{New York, NY, USA},
  \bibinfo{pages}{1957--1967}.
\newblock
\showISBNx{978-1-4503-3362-7}
\urldef\tempurl%
\url{https://doi.org/10.1145/2858036.2858134}
\showDOI{\tempurl}


\bibitem[Heun et~al\mbox{.}(2013)]%
        {heun_reality_2013}
\bibfield{author}{\bibinfo{person}{Valentin Heun}, \bibinfo{person}{James
  Hobin}, {and} \bibinfo{person}{Pattie Maes}.}
  \bibinfo{year}{2013}\natexlab{}.
\newblock \showarticletitle{Reality editor: programming smarter objects}. In
  \bibinfo{booktitle}{\emph{Proceedings of the 2013 {ACM} conference on
  {Pervasive} and ubiquitous computing adjunct publication}}
  \emph{(\bibinfo{series}{{UbiComp} '13 {Adjunct}})}.
  \bibinfo{publisher}{Association for Computing Machinery},
  \bibinfo{address}{New York, NY, USA}, \bibinfo{pages}{307--310}.
\newblock
\showISBNx{978-1-4503-2215-7}
\urldef\tempurl%
\url{https://doi.org/10.1145/2494091.2494185}
\showDOI{\tempurl}


\bibitem[Hirzle et~al\mbox{.}(2023)]%
        {hirzle_when_2023}
\bibfield{author}{\bibinfo{person}{Teresa Hirzle}, \bibinfo{person}{Florian
  Müller}, \bibinfo{person}{Fiona Draxler}, \bibinfo{person}{Martin Schmitz},
  \bibinfo{person}{Pascal Knierim}, {and} \bibinfo{person}{Kasper Hornbæk}.}
  \bibinfo{year}{2023}\natexlab{}.
\newblock \showarticletitle{When {XR} and {AI} {Meet} - {A} {Scoping} {Review}
  on {Extended} {Reality} and {Artificial} {Intelligence}}. In
  \bibinfo{booktitle}{\emph{Proceedings of the 2023 {CHI} {Conference} on
  {Human} {Factors} in {Computing} {Systems}}}. \bibinfo{publisher}{ACM},
  \bibinfo{address}{Hamburg Germany}, \bibinfo{pages}{1--45}.
\newblock
\showISBNx{978-1-4503-9421-5}
\urldef\tempurl%
\url{https://doi.org/10.1145/3544548.3581072}
\showDOI{\tempurl}


\bibitem[Inc(2024)]%
        {inc_arkit_2024}
\bibfield{author}{\bibinfo{person}{Apple Inc}.}
  \bibinfo{year}{2024}\natexlab{}.
\newblock \bibinfo{title}{{ARKit}}.
\newblock
\newblock
\urldef\tempurl%
\url{https://developer.apple.com/augmented-reality/arkit/}
\showURL{%
\tempurl}


\bibitem[Ishii(2008)]%
        {ishii_tangible_2008}
\bibfield{author}{\bibinfo{person}{Hiroshi Ishii}.}
  \bibinfo{year}{2008}\natexlab{}.
\newblock \showarticletitle{Tangible bits: beyond pixels}. In
  \bibinfo{booktitle}{\emph{Proceedings of the 2nd international conference on
  {Tangible} and embedded interaction}} \emph{(\bibinfo{series}{{TEI} '08})}.
  \bibinfo{publisher}{Association for Computing Machinery},
  \bibinfo{address}{New York, NY, USA}, \bibinfo{pages}{xv--xxv}.
\newblock
\showISBNx{978-1-60558-004-3}
\urldef\tempurl%
\url{https://doi.org/10.1145/1347390.1347392}
\showDOI{\tempurl}


\bibitem[Ishii and Ullmer(1997)]%
        {ishii_tangible_1997}
\bibfield{author}{\bibinfo{person}{Hiroshi Ishii} {and} \bibinfo{person}{Brygg
  Ullmer}.} \bibinfo{year}{1997}\natexlab{}.
\newblock \showarticletitle{Tangible bits: towards seamless interfaces between
  people, bits and atoms}. In \bibinfo{booktitle}{\emph{Proceedings of the
  {ACM} {SIGCHI} {Conference} on {Human} factors in computing systems}}
  \emph{(\bibinfo{series}{{CHI} '97})}. \bibinfo{publisher}{Association for
  Computing Machinery}, \bibinfo{address}{New York, NY, USA},
  \bibinfo{pages}{234--241}.
\newblock
\showISBNx{978-0-89791-802-2}
\urldef\tempurl%
\url{https://doi.org/10.1145/258549.258715}
\showDOI{\tempurl}


\bibitem[Jain et~al\mbox{.}(2023)]%
        {jain_ubi-touch_2023}
\bibfield{author}{\bibinfo{person}{Rahul Jain}, \bibinfo{person}{Jingyu Shi},
  \bibinfo{person}{Runlin Duan}, \bibinfo{person}{Zhengzhe Zhu},
  \bibinfo{person}{Xun Qian}, {and} \bibinfo{person}{Karthik Ramani}.}
  \bibinfo{year}{2023}\natexlab{}.
\newblock \showarticletitle{Ubi-{TOUCH}: {Ubiquitous} {Tangible} {Object}
  {Utilization} through {Consistent} {Hand}-object interaction in {Augmented}
  {Reality}}. In \bibinfo{booktitle}{\emph{Proceedings of the 36th {Annual}
  {ACM} {Symposium} on {User} {Interface} {Software} and {Technology}}}
  \emph{(\bibinfo{series}{{UIST} '23})}. \bibinfo{publisher}{Association for
  Computing Machinery}, \bibinfo{address}{New York, NY, USA},
  \bibinfo{pages}{1--18}.
\newblock
\showISBNx{9798400701320}
\urldef\tempurl%
\url{https://doi.org/10.1145/3586183.3606793}
\showDOI{\tempurl}


\bibitem[Kerbl et~al\mbox{.}(2023)]%
        {kerbl_3d_2023}
\bibfield{author}{\bibinfo{person}{Bernhard Kerbl}, \bibinfo{person}{Georgios
  Kopanas}, \bibinfo{person}{Thomas Leimkühler}, {and} \bibinfo{person}{George
  Drettakis}.} \bibinfo{year}{2023}\natexlab{}.
\newblock \showarticletitle{3d gaussian splatting for real-time radiance field
  rendering}.
\newblock \bibinfo{journal}{\emph{ACM Transactions on Graphics}}
  \bibinfo{volume}{42}, \bibinfo{number}{4} (\bibinfo{year}{2023}),
  \bibinfo{pages}{1--14}.
\newblock
\newblock
\shownote{Publisher: ACM}.


\bibitem[King et~al\mbox{.}(2018)]%
        {king_statistical_2018}
\bibfield{author}{\bibinfo{person}{Bruce~M King}, \bibinfo{person}{Patrick~J
  Rosopa}, {and} \bibinfo{person}{Edward~W Minium}.}
  \bibinfo{year}{2018}\natexlab{}.
\newblock \bibinfo{booktitle}{\emph{Statistical reasoning in the behavioral
  sciences}}.
\newblock \bibinfo{publisher}{John Wiley \& Sons}.
\newblock


\bibitem[Kudo et~al\mbox{.}(2021)]%
        {kudo_towards_2021}
\bibfield{author}{\bibinfo{person}{Yoshiki Kudo}, \bibinfo{person}{Anthony
  Tang}, \bibinfo{person}{Kazuyuki Fujita}, \bibinfo{person}{Isamu Endo},
  \bibinfo{person}{Kazuki Takashima}, {and} \bibinfo{person}{Yoshifumi
  Kitamura}.} \bibinfo{year}{2021}\natexlab{}.
\newblock \showarticletitle{Towards balancing {VR} immersion and bystander
  awareness}.
\newblock \bibinfo{journal}{\emph{Proceedings of the ACM on Human-Computer
  Interaction}} \bibinfo{volume}{5}, \bibinfo{number}{ISS}
  (\bibinfo{year}{2021}), \bibinfo{pages}{1--22}.
\newblock
\newblock
\shownote{Publisher: ACM New York, NY, USA}.


\bibitem[Lee et~al\mbox{.}(2024)]%
        {lee_gazepointar_2024}
\bibfield{author}{\bibinfo{person}{Jaewook Lee}, \bibinfo{person}{Jun Wang},
  \bibinfo{person}{Elizabeth Brown}, \bibinfo{person}{Liam Chu},
  \bibinfo{person}{Sebastian~S. Rodriguez}, {and} \bibinfo{person}{Jon~E.
  Froehlich}.} \bibinfo{year}{2024}\natexlab{}.
\newblock \showarticletitle{{GazePointAR}: {A} {Context}-{Aware} {Multimodal}
  {Voice} {Assistant} for {Pronoun} {Disambiguation} in {Wearable} {Augmented}
  {Reality}}. In \bibinfo{booktitle}{\emph{Proceedings of the {CHI}
  {Conference} on {Human} {Factors} in {Computing} {Systems}}}.
  \bibinfo{publisher}{ACM}, \bibinfo{address}{Honolulu HI USA},
  \bibinfo{pages}{1--20}.
\newblock
\showISBNx{9798400703300}
\urldef\tempurl%
\url{https://doi.org/10.1145/3613904.3642230}
\showDOI{\tempurl}


\bibitem[Lee et~al\mbox{.}(2022)]%
        {lee_evaluating_2022}
\bibfield{author}{\bibinfo{person}{Mina Lee}, \bibinfo{person}{Megha
  Srivastava}, \bibinfo{person}{Amelia Hardy}, \bibinfo{person}{John
  Thickstun}, \bibinfo{person}{Esin Durmus}, \bibinfo{person}{Ashwin
  Paranjape}, \bibinfo{person}{Ines Gerard-Ursin}, \bibinfo{person}{Xiang~Lisa
  Li}, \bibinfo{person}{Faisal Ladhak}, \bibinfo{person}{Frieda Rong}, {and}
  \bibinfo{person}{{others}}.} \bibinfo{year}{2022}\natexlab{}.
\newblock \showarticletitle{Evaluating human-language model interaction}.
\newblock \bibinfo{journal}{\emph{arXiv preprint arXiv:2212.09746}}
  (\bibinfo{year}{2022}).
\newblock


\bibitem[Lepinski et~al\mbox{.}(2009)]%
        {lepinski_context_2009}
\bibfield{author}{\bibinfo{person}{Julian Lepinski}, \bibinfo{person}{Eric
  Akaoka}, {and} \bibinfo{person}{Roel Vertegaal}.}
  \bibinfo{year}{2009}\natexlab{}.
\newblock \showarticletitle{Context menus for the real world: the
  stick-anywhere computer}. In \bibinfo{booktitle}{\emph{{CHI} '09 {Extended}
  {Abstracts} on {Human} {Factors} in {Computing} {Systems}}}
  \emph{(\bibinfo{series}{{CHI} {EA} '09})}. \bibinfo{publisher}{Association
  for Computing Machinery}, \bibinfo{address}{New York, NY, USA},
  \bibinfo{pages}{3499--3500}.
\newblock
\showISBNx{978-1-60558-247-4}
\urldef\tempurl%
\url{https://doi.org/10.1145/1520340.1520511}
\showDOI{\tempurl}
\newblock
\shownote{event-place: Boston, MA, USA}.


\bibitem[Li et~al\mbox{.}(2024)]%
        {li_omniactions_2024}
\bibfield{author}{\bibinfo{person}{Jiahao~Nick Li}, \bibinfo{person}{Yan Xu},
  \bibinfo{person}{Tovi Grossman}, \bibinfo{person}{Stephanie Santosa}, {and}
  \bibinfo{person}{Michelle Li}.} \bibinfo{year}{2024}\natexlab{}.
\newblock \showarticletitle{{OmniActions}: {Predicting} {Digital} {Actions} in
  {Response} to {Real}-{World} {Multimodal} {Sensory} {Inputs} with {LLMs}}. In
  \bibinfo{booktitle}{\emph{Proceedings of the {CHI} {Conference} on {Human}
  {Factors} in {Computing} {Systems}}} \emph{(\bibinfo{series}{{CHI} '24})}.
  \bibinfo{publisher}{Association for Computing Machinery},
  \bibinfo{address}{New York, NY, USA}, \bibinfo{pages}{1--22}.
\newblock
\showISBNx{9798400703300}
\urldef\tempurl%
\url{https://doi.org/10.1145/3613904.3642068}
\showDOI{\tempurl}


\bibitem[Li et~al\mbox{.}(2019)]%
        {li_holodoc_2019}
\bibfield{author}{\bibinfo{person}{Zhen Li}, \bibinfo{person}{Michelle Annett},
  \bibinfo{person}{Ken Hinckley}, \bibinfo{person}{Karan Singh}, {and}
  \bibinfo{person}{Daniel Wigdor}.} \bibinfo{year}{2019}\natexlab{}.
\newblock \showarticletitle{{HoloDoc}: {Enabling} {Mixed} {Reality}
  {Workspaces} that {Harness} {Physical} and {Digital} {Content}}. In
  \bibinfo{booktitle}{\emph{Proceedings of the 2019 {CHI} {Conference} on
  {Human} {Factors} in {Computing} {Systems}}} \emph{(\bibinfo{series}{{CHI}
  '19})}. \bibinfo{publisher}{Association for Computing Machinery},
  \bibinfo{address}{New York, NY, USA}, \bibinfo{pages}{1--14}.
\newblock
\showISBNx{978-1-4503-5970-2}
\urldef\tempurl%
\url{https://doi.org/10.1145/3290605.3300917}
\showDOI{\tempurl}
\newblock
\shownote{Issue: Paper 687 event-place: Glasgow, Scotland Uk}.


\bibitem[Lin et~al\mbox{.}(2015)]%
        {lin_microsoft_2015}
\bibfield{author}{\bibinfo{person}{Tsung-Yi Lin}, \bibinfo{person}{Michael
  Maire}, \bibinfo{person}{Serge Belongie}, \bibinfo{person}{Lubomir Bourdev},
  \bibinfo{person}{Ross Girshick}, \bibinfo{person}{James Hays},
  \bibinfo{person}{Pietro Perona}, \bibinfo{person}{Deva Ramanan},
  \bibinfo{person}{C.~Lawrence Zitnick}, {and} \bibinfo{person}{Piotr
  Dollár}.} \bibinfo{year}{2015}\natexlab{}.
\newblock \bibinfo{title}{Microsoft {COCO}: {Common} {Objects} in {Context}}.
\newblock
\newblock
\urldef\tempurl%
\url{http://arxiv.org/abs/1405.0312}
\showURL{%
\tempurl}
\newblock
\shownote{arXiv:1405.0312 [cs]}.


\bibitem[Lindlbauer et~al\mbox{.}(2019)]%
        {lindlbauer_context-aware_2019}
\bibfield{author}{\bibinfo{person}{David Lindlbauer},
  \bibinfo{person}{Anna~Maria Feit}, {and} \bibinfo{person}{Otmar Hilliges}.}
  \bibinfo{year}{2019}\natexlab{}.
\newblock \showarticletitle{Context-{Aware} {Online} {Adaptation} of {Mixed}
  {Reality} {Interfaces}}. In \bibinfo{booktitle}{\emph{Proceedings of the 32nd
  {Annual} {ACM} {Symposium} on {User} {Interface} {Software} and
  {Technology}}} \emph{(\bibinfo{series}{{UIST} '19})}.
  \bibinfo{publisher}{Association for Computing Machinery},
  \bibinfo{address}{New York, NY, USA}, \bibinfo{pages}{147--160}.
\newblock
\showISBNx{978-1-4503-6816-2}
\urldef\tempurl%
\url{https://doi.org/10.1145/3332165.3347945}
\showDOI{\tempurl}


\bibitem[Lindlbauer et~al\mbox{.}(2016)]%
        {lindlbauer_combining_2016}
\bibfield{author}{\bibinfo{person}{David Lindlbauer},
  \bibinfo{person}{Jens~Emil Grønbæk}, \bibinfo{person}{Morten Birk},
  \bibinfo{person}{Kim Halskov}, \bibinfo{person}{Marc Alexa}, {and}
  \bibinfo{person}{Jörg Müller}.} \bibinfo{year}{2016}\natexlab{}.
\newblock \showarticletitle{Combining {Shape}-{Changing} {Interfaces} and
  {Spatial} {Augmented} {Reality} {Enables} {Extended} {Object} {Appearance}}.
  In \bibinfo{booktitle}{\emph{Proceedings of the 2016 {CHI} {Conference} on
  {Human} {Factors} in {Computing} {Systems}}}. \bibinfo{publisher}{ACM},
  \bibinfo{address}{San Jose California USA}, \bibinfo{pages}{791--802}.
\newblock
\showISBNx{978-1-4503-3362-7}
\urldef\tempurl%
\url{https://doi.org/10.1145/2858036.2858457}
\showDOI{\tempurl}


\bibitem[Lindlbauer and Wilson(2018)]%
        {lindlbauer_remixed_2018}
\bibfield{author}{\bibinfo{person}{David Lindlbauer} {and}
  \bibinfo{person}{Andy~D Wilson}.} \bibinfo{year}{2018}\natexlab{}.
\newblock \showarticletitle{Remixed reality: {Manipulating} space and time in
  augmented reality}. In \bibinfo{booktitle}{\emph{Proceedings of the 2018
  {CHI} {Conference} on {Human} {Factors} in {Computing} {Systems}}}.
  \bibinfo{pages}{1--13}.
\newblock


\bibitem[Lu and Xu(2022)]%
        {lu_exploring_2022}
\bibfield{author}{\bibinfo{person}{Feiyu Lu} {and} \bibinfo{person}{Yan Xu}.}
  \bibinfo{year}{2022}\natexlab{}.
\newblock \showarticletitle{Exploring {Spatial} {UI} {Transition} {Mechanisms}
  with {Head}-{Worn} {Augmented} {Reality}}. In
  \bibinfo{booktitle}{\emph{Proceedings of the 2022 {CHI} {Conference} on
  {Human} {Factors} in {Computing} {Systems}}} \emph{(\bibinfo{series}{{CHI}
  '22})}. \bibinfo{publisher}{Association for Computing Machinery},
  \bibinfo{address}{New York, NY, USA}, \bibinfo{pages}{1--16}.
\newblock
\showISBNx{978-1-4503-9157-3}
\urldef\tempurl%
\url{https://doi.org/10.1145/3491102.3517723}
\showDOI{\tempurl}
\newblock
\shownote{Issue: Article 550 event-place: New Orleans, LA, USA}.


\bibitem[Lugaresi et~al\mbox{.}(2019)]%
        {lugaresi_mediapipe_2019}
\bibfield{author}{\bibinfo{person}{Camillo Lugaresi}, \bibinfo{person}{Jiuqiang
  Tang}, \bibinfo{person}{Hadon Nash}, \bibinfo{person}{Chris McClanahan},
  \bibinfo{person}{Esha Uboweja}, \bibinfo{person}{Michael Hays},
  \bibinfo{person}{Fan Zhang}, \bibinfo{person}{Chuo-Ling Chang},
  \bibinfo{person}{Ming~Guang Yong}, \bibinfo{person}{Juhyun Lee}, {and}
  \bibinfo{person}{{others}}.} \bibinfo{year}{2019}\natexlab{}.
\newblock \showarticletitle{Mediapipe: {A} framework for building perception
  pipelines}.
\newblock \bibinfo{journal}{\emph{arXiv preprint arXiv:1906.08172}}
  (\bibinfo{year}{2019}).
\newblock


\bibitem[Mayer et~al\mbox{.}(2020)]%
        {mayer_enhancing_2020}
\bibfield{author}{\bibinfo{person}{Sven Mayer}, \bibinfo{person}{Gierad Laput},
  {and} \bibinfo{person}{Chris Harrison}.} \bibinfo{year}{2020}\natexlab{}.
\newblock \showarticletitle{Enhancing {Mobile} {Voice} {Assistants} with
  {WorldGaze}}. In \bibinfo{booktitle}{\emph{Proceedings of the 2020 {CHI}
  {Conference} on {Human} {Factors} in {Computing} {Systems}}}
  \emph{(\bibinfo{series}{{CHI} '20})}. \bibinfo{publisher}{Association for
  Computing Machinery}, \bibinfo{address}{New York, NY, USA},
  \bibinfo{pages}{1--10}.
\newblock
\showISBNx{978-1-4503-6708-0}
\urldef\tempurl%
\url{https://doi.org/10.1145/3313831.3376479}
\showDOI{\tempurl}


\bibitem[Monteiro et~al\mbox{.}(2023)]%
        {monteiro_teachable_2023}
\bibfield{author}{\bibinfo{person}{Kyzyl Monteiro}, \bibinfo{person}{Ritik
  Vatsal}, \bibinfo{person}{Neil Chulpongsatorn}, \bibinfo{person}{Aman
  Parnami}, {and} \bibinfo{person}{Ryo Suzuki}.}
  \bibinfo{year}{2023}\natexlab{}.
\newblock \showarticletitle{Teachable {Reality}: {Prototyping} {Tangible}
  {Augmented} {Reality} with {Everyday} {Objects} by {Leveraging} {Interactive}
  {Machine} {Teaching}}. In \bibinfo{booktitle}{\emph{Proceedings of the 2023
  {CHI} {Conference} on {Human} {Factors} in {Computing} {Systems}}}
  \emph{(\bibinfo{series}{{CHI} '23})}. \bibinfo{publisher}{Association for
  Computing Machinery}, \bibinfo{address}{New York, NY, USA},
  \bibinfo{pages}{1--15}.
\newblock
\showISBNx{978-1-4503-9421-5}
\urldef\tempurl%
\url{https://doi.org/10.1145/3544548.3581449}
\showDOI{\tempurl}


\bibitem[Morris et~al\mbox{.}(2024)]%
        {morris_levels_2024}
\bibfield{author}{\bibinfo{person}{Meredith~Ringel Morris},
  \bibinfo{person}{Jascha Sohl-dickstein}, \bibinfo{person}{Noah Fiedel},
  \bibinfo{person}{Tris Warkentin}, \bibinfo{person}{Allan Dafoe},
  \bibinfo{person}{Aleksandra Faust}, \bibinfo{person}{Clement Farabet}, {and}
  \bibinfo{person}{Shane Legg}.} \bibinfo{year}{2024}\natexlab{}.
\newblock \bibinfo{title}{Levels of {AGI}: {Operationalizing} {Progress} on the
  {Path} to {AGI}}.
\newblock
\newblock
\newblock
\shownote{\_eprint: 2311.02462}.


\bibitem[Nebeling et~al\mbox{.}(2024)]%
        {nebeling_designing_2024}
\bibfield{author}{\bibinfo{person}{Michael Nebeling}, \bibinfo{person}{Mika
  Oki}, \bibinfo{person}{Mirko Gelsomini}, \bibinfo{person}{Gillian~R Hayes},
  \bibinfo{person}{Mark Billinghurst}, \bibinfo{person}{Kenji Suzuki}, {and}
  \bibinfo{person}{Roland Graf}.} \bibinfo{year}{2024}\natexlab{}.
\newblock \showarticletitle{Designing {Inclusive} {Future} {Augmented}
  {Realities}}. In \bibinfo{booktitle}{\emph{Extended {Abstracts} of the 2024
  {CHI} {Conference} on {Human} {Factors} in {Computing} {Systems}}}
  \emph{(\bibinfo{series}{{CHI} {EA} '24})}. \bibinfo{publisher}{Association
  for Computing Machinery}, \bibinfo{address}{New York, NY, USA},
  \bibinfo{pages}{1--6}.
\newblock
\showISBNx{9798400703317}
\urldef\tempurl%
\url{https://doi.org/10.1145/3613905.3636313}
\showDOI{\tempurl}


\bibitem[Pourmemar and Poullis(2019)]%
        {pourmemar_visualizing_2019}
\bibfield{author}{\bibinfo{person}{Majid Pourmemar} {and}
  \bibinfo{person}{Charalambos Poullis}.} \bibinfo{year}{2019}\natexlab{}.
\newblock \showarticletitle{Visualizing and {Interacting} with {Hierarchical}
  {Menus} in {Immersive} {Augmented} {Reality}}. In
  \bibinfo{booktitle}{\emph{Proceedings of the 17th {International}
  {Conference} on {Virtual}-{Reality} {Continuum} and its {Applications} in
  {Industry}}} \emph{(\bibinfo{series}{{VRCAI} '19})}.
  \bibinfo{publisher}{Association for Computing Machinery},
  \bibinfo{address}{New York, NY, USA}, \bibinfo{pages}{1--9}.
\newblock
\showISBNx{978-1-4503-7002-8}
\urldef\tempurl%
\url{https://doi.org/10.1145/3359997.3365693}
\showDOI{\tempurl}


\bibitem[Rajaram and Nebeling(2022)]%
        {rajaram_paper_2022}
\bibfield{author}{\bibinfo{person}{Shwetha Rajaram} {and}
  \bibinfo{person}{Michael Nebeling}.} \bibinfo{year}{2022}\natexlab{}.
\newblock \showarticletitle{Paper {Trail}: {An} {Immersive} {Authoring}
  {System} for {Augmented} {Reality} {Instructional} {Experiences}}. In
  \bibinfo{booktitle}{\emph{Proceedings of the 2022 {CHI} {Conference} on
  {Human} {Factors} in {Computing} {Systems}}} \emph{(\bibinfo{series}{{CHI}
  '22})}. \bibinfo{publisher}{Association for Computing Machinery},
  \bibinfo{address}{New York, NY, USA}, \bibinfo{pages}{1--16}.
\newblock
\showISBNx{978-1-4503-9157-3}
\urldef\tempurl%
\url{https://doi.org/10.1145/3491102.3517486}
\showDOI{\tempurl}


\bibitem[Rogers et~al\mbox{.}(2020)]%
        {rogers_understanding_2020}
\bibfield{author}{\bibinfo{person}{Wendy~A Rogers}, \bibinfo{person}{Tracy~L
  Mitzner}, {and} \bibinfo{person}{Michael~T Bixter}.}
  \bibinfo{year}{2020}\natexlab{}.
\newblock \showarticletitle{Understanding the potential of technology to
  support enhanced activities of daily living ({EADLs}).}
\newblock \bibinfo{journal}{\emph{Gerontechnology}} \bibinfo{volume}{19},
  \bibinfo{number}{2} (\bibinfo{year}{2020}).
\newblock


\bibitem[Samp and Decker(2010)]%
        {samp_supporting_2010}
\bibfield{author}{\bibinfo{person}{Krystian Samp} {and} \bibinfo{person}{Stefan
  Decker}.} \bibinfo{year}{2010}\natexlab{}.
\newblock \showarticletitle{Supporting menu design with radial layouts}. In
  \bibinfo{booktitle}{\emph{Proceedings of the {International} {Conference} on
  {Advanced} {Visual} {Interfaces}}}. \bibinfo{publisher}{ACM},
  \bibinfo{address}{Roma Italy}, \bibinfo{pages}{155--162}.
\newblock
\showISBNx{978-1-4503-0076-6}
\urldef\tempurl%
\url{https://doi.org/10.1145/1842993.1843021}
\showDOI{\tempurl}


\bibitem[Speicher et~al\mbox{.}(2019)]%
        {speicher_what_2019}
\bibfield{author}{\bibinfo{person}{Maximilian Speicher},
  \bibinfo{person}{Brian~D. Hall}, {and} \bibinfo{person}{Michael Nebeling}.}
  \bibinfo{year}{2019}\natexlab{}.
\newblock \showarticletitle{What is {Mixed} {Reality}?}. In
  \bibinfo{booktitle}{\emph{Proceedings of the 2019 {CHI} {Conference} on
  {Human} {Factors} in {Computing} {Systems}}}. \bibinfo{publisher}{ACM},
  \bibinfo{address}{Glasgow Scotland Uk}, \bibinfo{pages}{1--15}.
\newblock
\showISBNx{978-1-4503-5970-2}
\urldef\tempurl%
\url{https://doi.org/10.1145/3290605.3300767}
\showDOI{\tempurl}


\bibitem[Strecker et~al\mbox{.}(2023)]%
        {strecker_mr_2023}
\bibfield{author}{\bibinfo{person}{Jannis Strecker}, \bibinfo{person}{Khakim
  Akhunov}, \bibinfo{person}{Federico Carbone}, \bibinfo{person}{Kimberly
  García}, \bibinfo{person}{Kenan Bektaş}, \bibinfo{person}{Andres Gomez},
  \bibinfo{person}{Simon Mayer}, {and} \bibinfo{person}{Kasim~Sinan Yildirim}.}
  \bibinfo{year}{2023}\natexlab{}.
\newblock \showarticletitle{{MR} {Object} {Identification} and {Interaction}:
  {Fusing} {Object} {Situation} {Information} from {Heterogeneous} {Sources}}.
\newblock \bibinfo{journal}{\emph{Proceedings of the ACM on Interactive,
  Mobile, Wearable and Ubiquitous Technologies}} \bibinfo{volume}{7},
  \bibinfo{number}{3} (\bibinfo{date}{Sept.} \bibinfo{year}{2023}),
  \bibinfo{pages}{1--26}.
\newblock
\showISSN{2474-9567}
\urldef\tempurl%
\url{https://doi.org/10.1145/3610879}
\showDOI{\tempurl}


\bibitem[Suh et~al\mbox{.}(2024)]%
        {suh_luminate_2024}
\bibfield{author}{\bibinfo{person}{Sangho Suh}, \bibinfo{person}{Meng Chen},
  \bibinfo{person}{Bryan Min}, \bibinfo{person}{Toby Jia-Jun Li}, {and}
  \bibinfo{person}{Haijun Xia}.} \bibinfo{year}{2024}\natexlab{}.
\newblock \bibinfo{title}{Luminate: {Structured} {Generation} and {Exploration}
  of {Design} {Space} with {Large} {Language} {Models} for {Human}-{AI}
  {Co}-{Creation}}.
\newblock
\newblock
\urldef\tempurl%
\url{https://doi.org/10.1145/3613904.3642400}
\showDOI{\tempurl}
\newblock
\shownote{arXiv:2310.12953 [cs]}.


\bibitem[Suzuki et~al\mbox{.}(2023)]%
        {suzuki_xr_2023}
\bibfield{author}{\bibinfo{person}{Ryo Suzuki}, \bibinfo{person}{Mar
  Gonzalez-Franco}, \bibinfo{person}{Misha Sra}, {and} \bibinfo{person}{David
  Lindlbauer}.} \bibinfo{year}{2023}\natexlab{}.
\newblock \showarticletitle{{XR} and {AI}: {AI}-{Enabled} {Virtual},
  {Augmented}, and {Mixed} {Reality}}. In \bibinfo{booktitle}{\emph{Adjunct
  {Proceedings} of the 36th {Annual} {ACM} {Symposium} on {User} {Interface}
  {Software} and {Technology}}}. \bibinfo{publisher}{ACM},
  \bibinfo{address}{San Francisco CA USA}, \bibinfo{pages}{1--3}.
\newblock
\showISBNx{9798400700965}
\urldef\tempurl%
\url{https://doi.org/10.1145/3586182.3617432}
\showDOI{\tempurl}


\bibitem[Suzuki et~al\mbox{.}(2020)]%
        {suzuki_realitysketch_2020}
\bibfield{author}{\bibinfo{person}{Ryo Suzuki}, \bibinfo{person}{Rubaiat~Habib
  Kazi}, \bibinfo{person}{Li-yi Wei}, \bibinfo{person}{Stephen DiVerdi},
  \bibinfo{person}{Wilmot Li}, {and} \bibinfo{person}{Daniel Leithinger}.}
  \bibinfo{year}{2020}\natexlab{}.
\newblock \showarticletitle{{RealitySketch}: {Embedding} {Responsive}
  {Graphics} and {Visualizations} in {AR} through {Dynamic} {Sketching}}. In
  \bibinfo{booktitle}{\emph{Proceedings of the 33rd {Annual} {ACM} {Symposium}
  on {User} {Interface} {Software} and {Technology}}}.
  \bibinfo{publisher}{ACM}, \bibinfo{address}{Virtual Event USA},
  \bibinfo{pages}{166--181}.
\newblock
\showISBNx{978-1-4503-7514-6}
\urldef\tempurl%
\url{https://doi.org/10.1145/3379337.3415892}
\showDOI{\tempurl}


\bibitem[Tian et~al\mbox{.}(2023)]%
        {tian_diffuse_2023}
\bibfield{author}{\bibinfo{person}{Junjiao Tian}, \bibinfo{person}{Lavisha
  Aggarwal}, \bibinfo{person}{Andrea Colaco}, \bibinfo{person}{Zsolt Kira},
  {and} \bibinfo{person}{Mar Gonzalez-Franco}.}
  \bibinfo{year}{2023}\natexlab{}.
\newblock \showarticletitle{Diffuse, attend, and segment: {Unsupervised}
  zero-shot segmentation using stable diffusion}.
\newblock \bibinfo{journal}{\emph{arXiv preprint arXiv:2308.12469}}
  (\bibinfo{year}{2023}).
\newblock


\bibitem[Tidwell et~al\mbox{.}(2020)]%
        {tidwell_designing_2020}
\bibfield{author}{\bibinfo{person}{Jenifer Tidwell}, \bibinfo{person}{Charles
  Brewer}, {and} \bibinfo{person}{Aynne Valencia}.}
  \bibinfo{year}{2020}\natexlab{}.
\newblock \bibinfo{booktitle}{\emph{Designing {Interfaces}: {Patterns} for
  {Effective} {Interaction} {Design}} (\bibinfo{edition}{3} ed.)}.
\newblock \bibinfo{publisher}{O'Reilly Media}.
\newblock
\showISBNx{978-1-4920-5196-1}
\urldef\tempurl%
\url{https://www.amazon.com/Designing-Interfaces-Patterns-Effective-Interaction/dp/1492051969}
\showURL{%
\tempurl}


\bibitem[Tseng(2023)]%
        {tseng_understanding_2023}
\bibfield{author}{\bibinfo{person}{Wen-Jie Tseng}.}
  \bibinfo{year}{2023}\natexlab{}.
\newblock \showarticletitle{Understanding {Physical} {Breakdowns} in {Virtual}
  {Reality}}. In \bibinfo{booktitle}{\emph{Extended {Abstracts} of the 2023
  {CHI} {Conference} on {Human} {Factors} in {Computing} {Systems}}}
  \emph{(\bibinfo{series}{{CHI} {EA} '23})}. \bibinfo{publisher}{Association
  for Computing Machinery}, \bibinfo{address}{New York, NY, USA}.
\newblock
\showISBNx{978-1-4503-9422-2}
\newblock
\shownote{event-place: Hamburg,Germany}.


\bibitem[Tseng et~al\mbox{.}(2022)]%
        {tseng_dark_2022}
\bibfield{author}{\bibinfo{person}{Wen-Jie Tseng}, \bibinfo{person}{Elise
  Bonnail}, \bibinfo{person}{Mark McGill}, \bibinfo{person}{Mohamed Khamis},
  \bibinfo{person}{Eric Lecolinet}, \bibinfo{person}{Samuel Huron}, {and}
  \bibinfo{person}{Jan Gugenheimer}.} \bibinfo{year}{2022}\natexlab{}.
\newblock \showarticletitle{The {Dark} {Side} of {Perceptual} {Manipulations}
  in {Virtual} {Reality}}. In \bibinfo{booktitle}{\emph{Proceedings of the 2022
  {CHI} {Conference} on {Human} {Factors} in {Computing} {Systems}}}
  \emph{(\bibinfo{series}{{CHI} '22})}. \bibinfo{publisher}{Association for
  Computing Machinery}, \bibinfo{address}{New York, NY, USA}.
\newblock
\showISBNx{978-1-4503-9157-3}
\newblock
\shownote{event-place: New Orleans, LA,USA}.


\bibitem[Valentin et~al\mbox{.}(2018)]%
        {valentin_depth_2018}
\bibfield{author}{\bibinfo{person}{Julien Valentin}, \bibinfo{person}{Adarsh
  Kowdle}, \bibinfo{person}{Jonathan~T Barron}, \bibinfo{person}{Neal Wadhwa},
  \bibinfo{person}{Max Dzitsiuk}, \bibinfo{person}{Michael Schoenberg},
  \bibinfo{person}{Vivek Verma}, \bibinfo{person}{Ambrus Csaszar},
  \bibinfo{person}{Eric Turner}, \bibinfo{person}{Ivan Dryanovski}, {and}
  \bibinfo{person}{{others}}.} \bibinfo{year}{2018}\natexlab{}.
\newblock \showarticletitle{Depth from motion for smartphone {AR}}.
\newblock \bibinfo{journal}{\emph{ACM Transactions on Graphics (ToG)}}
  \bibinfo{volume}{37}, \bibinfo{number}{6} (\bibinfo{year}{2018}),
  \bibinfo{pages}{1--19}.
\newblock
\newblock
\shownote{Publisher: ACM New York, NY, USA}.


\bibitem[Wang and Zhang(2024)]%
        {wang_texturesight_2024}
\bibfield{author}{\bibinfo{person}{Xue Wang} {and} \bibinfo{person}{Yang
  Zhang}.} \bibinfo{year}{2024}\natexlab{}.
\newblock \showarticletitle{{TextureSight}: {Texture} {Detection} for {Routine}
  {Activity} {Awareness} with {Wearable} {Laser} {Speckle} {Imaging}}.
\newblock \bibinfo{journal}{\emph{Proc. ACM Interact. Mob. Wearable Ubiquitous
  Technol.}} \bibinfo{volume}{7}, \bibinfo{number}{4} (\bibinfo{date}{Jan.}
  \bibinfo{year}{2024}), \bibinfo{pages}{184:1--184:27}.
\newblock
\urldef\tempurl%
\url{https://doi.org/10.1145/3631413}
\showDOI{\tempurl}


\bibitem[Wu et~al\mbox{.}(2017)]%
        {wu_hickhyman_2017}
\bibfield{author}{\bibinfo{person}{Tingting Wu}, \bibinfo{person}{Alexander~J
  Dufford}, \bibinfo{person}{Laura~J Egan}, \bibinfo{person}{Melissa-Ann
  Mackie}, \bibinfo{person}{Cong Chen}, \bibinfo{person}{Changhe Yuan},
  \bibinfo{person}{Chao Chen}, \bibinfo{person}{Xiaobo Li},
  \bibinfo{person}{Xun Liu}, \bibinfo{person}{Patrick~R Hof}, {and}
  \bibinfo{person}{Jin Fan}.} \bibinfo{year}{2017}\natexlab{}.
\newblock \showarticletitle{Hick–{Hyman} {Law} is {Mediated} by the
  {Cognitive} {Control} {Network} in the {Brain}}.
\newblock \bibinfo{journal}{\emph{Cerebral Cortex}} \bibinfo{volume}{28},
  \bibinfo{number}{7} (\bibinfo{date}{May} \bibinfo{year}{2017}),
  \bibinfo{pages}{2267--2282}.
\newblock
\showISSN{1047-3211}


\bibitem[Xiang et~al\mbox{.}(2023)]%
        {xiang_deep_2023}
\bibfield{author}{\bibinfo{person}{Hanyu Xiang}, \bibinfo{person}{Qin Zou},
  \bibinfo{person}{Muhammad~Ali Nawaz}, \bibinfo{person}{Xianfeng Huang},
  \bibinfo{person}{Fan Zhang}, {and} \bibinfo{person}{Hongkai Yu}.}
  \bibinfo{year}{2023}\natexlab{}.
\newblock \showarticletitle{Deep learning for image inpainting: {A} survey}.
\newblock \bibinfo{journal}{\emph{Pattern Recognition}}  \bibinfo{volume}{134}
  (\bibinfo{year}{2023}), \bibinfo{pages}{109046}.
\newblock
\newblock
\shownote{Publisher: Elsevier}.


\bibitem[Xiao et~al\mbox{.}(2022)]%
        {xiao_imarker_2022}
\bibfield{author}{\bibinfo{person}{Chang Xiao}, \bibinfo{person}{Ryan Rossi},
  {and} \bibinfo{person}{Eunyee Koh}.} \bibinfo{year}{2022}\natexlab{}.
\newblock \showarticletitle{{iMarker}: {Instant} and {True}-to-scale {AR} with
  {Invisible} {Markers}}. In \bibinfo{booktitle}{\emph{Adjunct {Proceedings} of
  the 35th {Annual} {ACM} {Symposium} on {User} {Interface} {Software} and
  {Technology}}} \emph{(\bibinfo{series}{{UIST} '22 {Adjunct}})}.
  \bibinfo{publisher}{Association for Computing Machinery},
  \bibinfo{address}{New York, NY, USA}, \bibinfo{pages}{1--3}.
\newblock
\showISBNx{978-1-4503-9321-8}
\urldef\tempurl%
\url{https://doi.org/10.1145/3526114.3558721}
\showDOI{\tempurl}


\bibitem[Xu et~al\mbox{.}(2022)]%
        {xu_arshopping_2022}
\bibfield{author}{\bibinfo{person}{Bingjie Xu}, \bibinfo{person}{Shunan Guo},
  \bibinfo{person}{Eunyee Koh}, \bibinfo{person}{Jane Hoffswell},
  \bibinfo{person}{Ryan Rossi}, {and} \bibinfo{person}{Fan Du}.}
  \bibinfo{year}{2022}\natexlab{}.
\newblock \showarticletitle{{ARShopping}: {In}-{Store} {Shopping} {Decision}
  {Support} {Through} {Augmented} {Reality} and {Immersive} {Visualization}}.
  In \bibinfo{booktitle}{\emph{2022 {IEEE} {Visualization} and {Visual}
  {Analytics} ({VIS})}}. \bibinfo{pages}{120--124}.
\newblock
\urldef\tempurl%
\url{https://doi.org/10.1109/VIS54862.2022.00033}
\showDOI{\tempurl}
\newblock
\shownote{ISSN: 2771-9553}.


\bibitem[Xu et~al\mbox{.}(2023)]%
        {xu_xair_2023}
\bibfield{author}{\bibinfo{person}{Xuhai Xu}, \bibinfo{person}{Anna Yu},
  \bibinfo{person}{Tanya~R. Jonker}, \bibinfo{person}{Kashyap Todi},
  \bibinfo{person}{Feiyu Lu}, \bibinfo{person}{Xun Qian},
  \bibinfo{person}{João~Marcelo Evangelista~Belo}, \bibinfo{person}{Tianyi
  Wang}, \bibinfo{person}{Michelle Li}, \bibinfo{person}{Aran Mun},
  \bibinfo{person}{Te-Yen Wu}, \bibinfo{person}{Junxiao Shen},
  \bibinfo{person}{Ting Zhang}, \bibinfo{person}{Narine Kokhlikyan},
  \bibinfo{person}{Fulton Wang}, \bibinfo{person}{Paul Sorenson},
  \bibinfo{person}{Sophie Kim}, {and} \bibinfo{person}{Hrvoje Benko}.}
  \bibinfo{year}{2023}\natexlab{}.
\newblock \showarticletitle{{XAIR}: {A} {Framework} of {Explainable} {AI} in
  {Augmented} {Reality}}. In \bibinfo{booktitle}{\emph{Proceedings of the 2023
  {CHI} {Conference} on {Human} {Factors} in {Computing} {Systems}}}.
  \bibinfo{publisher}{ACM}, \bibinfo{address}{Hamburg Germany},
  \bibinfo{pages}{1--30}.
\newblock
\showISBNx{978-1-4503-9421-5}
\urldef\tempurl%
\url{https://doi.org/10.1145/3544548.3581500}
\showDOI{\tempurl}


\bibitem[Yang and Landay(2019)]%
        {yang_infoled_2019}
\bibfield{author}{\bibinfo{person}{Jackie~(Junrui) Yang} {and}
  \bibinfo{person}{James~A. Landay}.} \bibinfo{year}{2019}\natexlab{}.
\newblock \showarticletitle{{InfoLED}: {Augmenting} {LED} {Indicator} {Lights}
  for {Device} {Positioning} and {Communication}}. In
  \bibinfo{booktitle}{\emph{Proceedings of the 32nd {Annual} {ACM} {Symposium}
  on {User} {Interface} {Software} and {Technology}}}
  \emph{(\bibinfo{series}{{UIST} '19})}. \bibinfo{publisher}{Association for
  Computing Machinery}, \bibinfo{address}{New York, NY, USA},
  \bibinfo{pages}{175--187}.
\newblock
\showISBNx{978-1-4503-6816-2}
\urldef\tempurl%
\url{https://doi.org/10.1145/3332165.3347954}
\showDOI{\tempurl}


\bibitem[Yin et~al\mbox{.}(2024)]%
        {yin_survey_2024}
\bibfield{author}{\bibinfo{person}{Shukang Yin}, \bibinfo{person}{Chaoyou Fu},
  \bibinfo{person}{Sirui Zhao}, \bibinfo{person}{Ke Li}, \bibinfo{person}{Xing
  Sun}, \bibinfo{person}{Tong Xu}, {and} \bibinfo{person}{Enhong Chen}.}
  \bibinfo{year}{2024}\natexlab{}.
\newblock \bibinfo{title}{A {Survey} on {Multimodal} {Large} {Language}
  {Models}}.
\newblock
\newblock
\newblock
\shownote{\_eprint: 2306.13549}.


\bibitem[Zeng and Zhang(2014)]%
        {zeng_multiple_2014}
\bibfield{author}{\bibinfo{person}{Yuguang Zeng} {and}
  \bibinfo{person}{Jingyuan Zhang}.} \bibinfo{year}{2014}\natexlab{}.
\newblock \showarticletitle{Multiple user context menus for large displays}. In
  \bibinfo{booktitle}{\emph{Proceedings of the 2014 {ACM} {Southeast}
  {Regional} {Conference}}} \emph{(\bibinfo{series}{{ACM} {SE} '14})}.
  \bibinfo{publisher}{Association for Computing Machinery},
  \bibinfo{address}{New York, NY, USA}, \bibinfo{pages}{1--4}.
\newblock
\showISBNx{978-1-4503-2923-1}
\urldef\tempurl%
\url{https://doi.org/10.1145/2638404.2638518}
\showDOI{\tempurl}
\newblock
\shownote{Issue: Article 44 event-place: Kennesaw, Georgia}.


\bibitem[Zhu et~al\mbox{.}(2022)]%
        {zhu_mecharspace_2022}
\bibfield{author}{\bibinfo{person}{Zhengzhe Zhu}, \bibinfo{person}{Ziyi Liu},
  \bibinfo{person}{Tianyi Wang}, \bibinfo{person}{Youyou Zhang},
  \bibinfo{person}{Xun Qian}, \bibinfo{person}{Pashin~Farsak Raja},
  \bibinfo{person}{Ana Villanueva}, {and} \bibinfo{person}{Karthik Ramani}.}
  \bibinfo{year}{2022}\natexlab{}.
\newblock \showarticletitle{{MechARspace}: {An} {Authoring} {System} {Enabling}
  {Bidirectional} {Binding} of {Augmented} {Reality} with {Toys} in
  {Real}-time}. In \bibinfo{booktitle}{\emph{Proceedings of the 35th {Annual}
  {ACM} {Symposium} on {User} {Interface} {Software} and {Technology}}}
  \emph{(\bibinfo{series}{{UIST} '22})}. \bibinfo{publisher}{Association for
  Computing Machinery}, \bibinfo{address}{New York, NY, USA},
  \bibinfo{pages}{1--16}.
\newblock
\showISBNx{978-1-4503-9320-1}
\urldef\tempurl%
\url{https://doi.org/10.1145/3526113.3545668}
\showDOI{\tempurl}


\end{thebibliography}

% %%
% %% If your work has an appendix, this is the place to put it.
\appendix

\section{Study Materials}
\subsection{HALIE Survey} \label{post-condition survey}
Following the completion of all tasks in a given condition (\systemName, Chatbot), participants rated their agreement with the following statements on a 5-point \textit{Likert} scale, and provided open ended answers to the questions denoted with (*O):

\begin{itemize}
    \setlength\itemsep{0.3em}
    
    \item [H] \textbf{(Helpfulness)} Independent of its fluency, the AI Tool was helpful for completing my task.  
    
    \item [*O1] \textbf{(Helpfulness)}  What kinds of aspects did you find helpful or not helpful and why? (Give a concrete example if possible.)
    
    \item [J] \textbf{(Enjoyment)} It was enjoyable using the AI tool to accomplish the tasks. 
    
    \item [S] \textbf{(Satisfaction)} Independent of its fluency, I am satisfied with \textit{how} the AI tool provided its answers.
    
    \item [R] \textbf{(Responsiveness)} Independent of its fluency, I found the AI tool to be a responsive system.
    
    \item [E] \textbf{(Ease)} Overall, it was easy to interact with the AI Tool and accomplish the tasks.  
    
    \item [ER] \textbf{(Ease)} Getting information about an object was easy using the AI tool.  
    
    \item [E2] \textbf{(Ease)} Comparing two objects was easy using the AI tool.   
    
    \item [EN] \textbf{(Ease)} Comparing more than two objects was easy using the AI tool. 
    
    \item [*O2] \textbf{(Change)} Did you change how you chose to interact with the AI Tool over the course of the task? If so, how?
    
    \item [*O3] \textbf{(Description)} What adjectives would you use to describe the AI Tool?
    
    \item [*O4] \textbf{(Compare)} How did this interaction compare to your regular in-person shopping and at-home task experiences?

\end{itemize}

\subsection{Form Factor Survey} \label{follow-up survey}
For each question, participants selected either Chatbot or \systemName. They did this questionnaire twice, once we were asking participants to think \systemName~ were going to run on a Phone, the second time thinking \systemName~ would run on a headset form factor. This questionnaire included the adapted HALIE questions (H, J, S, R, ER, E2, EN) as well as an additional set of questions detailed below:

\begin{itemize}
    \setlength\itemsep{0.3em}

    \item [EC] \textbf{(Ease Communications)} Sending a message about one of the grocery items would be easier on headset/phone using:
    
    \item [ET] \textbf{(Ease Timer)} Setting a timer would be easier on headset/phone using:
    
    \item [EN] \textbf{(Ease Note)} Creating a note (e.g., reminder to buy more juice) would be easier on a headset/phone using:
    
    \item [IS] \textbf{(Improved Shopping)} Which AI Tool running on a headset/phone would represent a better change compared to your current experience with (in-person) shopping?   
    
    \item [P] \textbf{(Preference)} Overall, which AI tool would you prefer on headset/phone? 
    
 \end{itemize}

\subsection{Results}

\begin{figure}[H]
  \centering
  \includegraphics[width=0.8\linewidth]{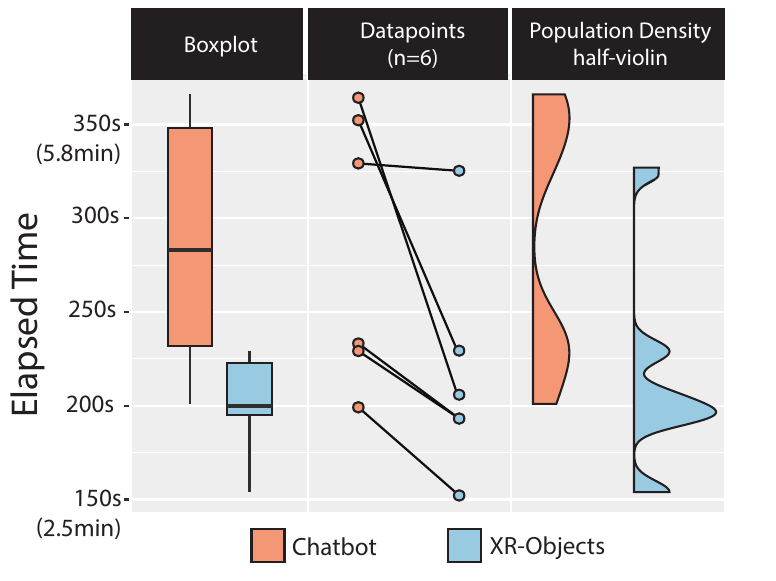}
  \caption{Comparison of task completion times between \systemName~ and Chatbot. The box plot visualizes the distribution of data, individual data points, within-subject comparisons (n=6), and the distribution for each condition.
%   Elapsed time to complete the tasks in the two different conditions. In this figure we show the box plot of the data, the individual data-points and within subjects comparison (n=6), and the corresponding distributions per each condition.
  }
  \Description{The boxplot shows that Chatbot took longer than XR-Objects. The individual datapoints, as well as population density half-violin plots are also shown for comparison.}
  \label{fig:time}
\end{figure}

\begin{figure}[H]
  \centering
  \includegraphics[width=0.8\linewidth]{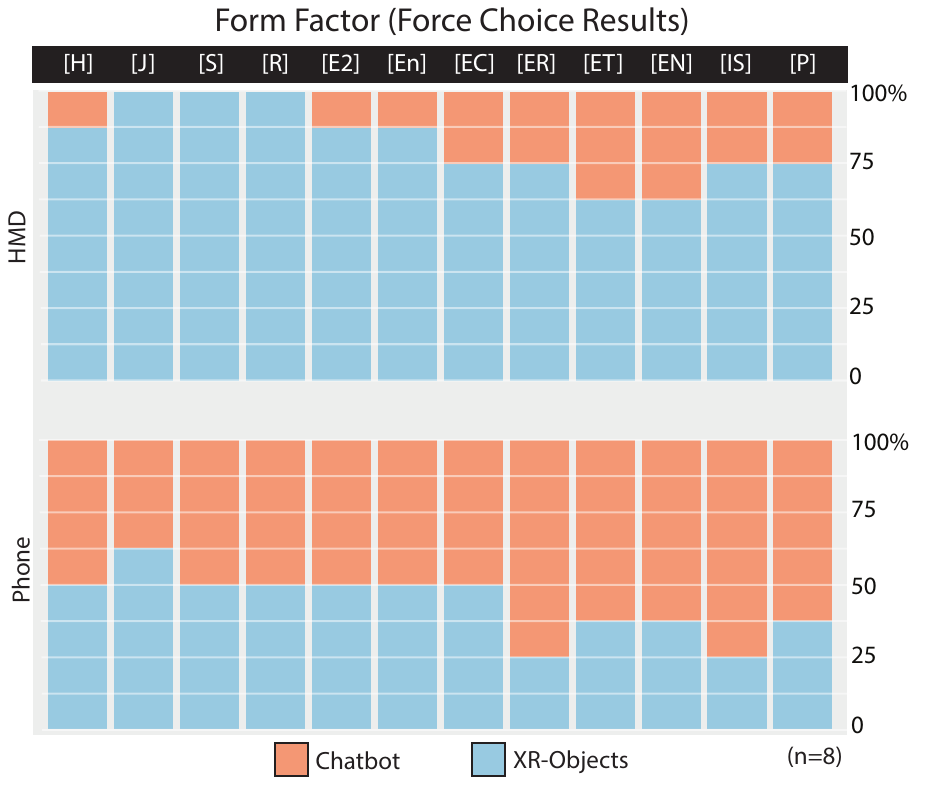}
  \caption{\textit{Marimekko}-chart mosaic with the form factor survey results, showing a preference for \systemName~ over Chatbot on the HMD condition.
  % This chart shows the proportion of respondents who selected XR Objects vs Chatbot for each form factor and measure in the post-study survey. Refer to Appendix X for complete list of measures.
  }
  \Description{The mosaic chart shows a preference for XR-Objects over Chatbot on the HMD condition, compared to the phone condition, where the participants were indifferent across most of the conditions.}
  \label{fig:formfactor}
\end{figure}

The analysis of completion time and form factory survey is depicted in Figure~\ref{fig:time} and Figure~\ref{fig:formfactor}, respectively.

\end{document}